# Kinetic energy densities based on the fourth order gradient expansion: performance in different classes of materials and improvement via machine learning

*Pavlo Golub, Sergei Manzhos[1]*

Department of Mechanical Engineering, National University of Singapore, Block EA #07-08, 9 Engineering Drive 1, Singapore 117576.

## Abstract

We study the performance of fourth-order gradient expansions of the kinetic energy density (KED) in semi-local kinetic energy functionals depending on the density-dependent variables. The formal fourth-order expansion is convergent for periodic systems and small molecules but does not improve over the second-order expansion (Thomas-Fermi term plus one-ninth of von Weizsäcker term). Linear fitting of the expansion coefficients somewhat improves on the formal expansion. The tuning of the fourth order expansion coefficients allows for better reproducibility of Kohn-Sham kinetic energy density than the tuning of the second-order expansion coefficients alone. The possibility of a much more accurate match with the Kohn-Sham kinetic energy density by using neural networks trained using the terms of the 4$^{th}$ order expansion as density-dependent variables is demonstrated. We obtain ultra-low fitting errors without overfitting. Small single hidden layer neural networks can provide good accuracy in separate KED fits of each compound, while for joint fitting of KEDs of multiple compounds multiple hidden layers were required to achieve good fit quality. The critical issue of data distribution is highlighted. We also show the critical role of pseudopotentials in the performance of the expansion, where in the case of a too rapid decay of the valence density at the nucleus with some pseudopotentials, numeric instabilities arise.

[1] Department of Mechanical Engineering, National University of Singapore, Block EA #07-08, 9 Engineering Drive 1, Singapore 117576. Tel: +65 6516 4605. Fax: +65 6779 1459. E-mail: mpemanzh@nus.edu.sg.

## Introduction

The search for the orbital-free form of non-interacting kinetic energy functional (KEF), or more narrowly kinetic energy density functional (KEDF), remains one of the most challenging tasks in density-functional theory (DFT).[1,2] Finding such an expression would be an enormous boon for computational material science, since in that case instead of solving self-consistently for a set of single-electron wavefunctions using the Kohn-Sham equation,[3] one would need to solve self-consistently only the minimization problem of the total energy of the system $E$ with respect to the electron density $n(\bm{r})$, $E[n(\bm{r})]$,

$$\frac{\partial E[n]}{\partial n(\bm{r})} = \mu - V_{ext}(\bm{r})$$

(1)

Here $\mu$ is the chemical potential that aims to ensure the normalization condition $\int n(\bm{r})d\bm{r} = N_{el}$, where $N_{el}$ is the number of electrons, and is $V_{ext}(\bm{r})$ external field (for simplicity, we will ignore spin). Indeed, while other components in DFT partitioning of the total energy – the Hartree energy, nuclear-electron interaction energy and exchange-correlation energy – can be explicitly represented via electron density with satisfactory accuracy, good approximations to the kinetic energy (KE) or kinetic energy density (KED) as a functional of only density-dependent quantities are still missing. They are the main subject of developments in orbital-free DFT (OF-DFT).[4] Existing approximations to the KEF work well, i.e. provide accuracy similar to KS-DFT, only when the density is slowly varying, specifically, for light metals,[5-10] and some simplest semiconductors,[11] but there is still no KEF that would be sufficiently accurate on molecular densities to allow for molecular structure optimization with accuracy approaching that of Kohn-Sham DFT in applications.

One of the earliest and most straightforward methods for representing the KE is an expansion in different orders of the gradient of the electron density. Initially introduced by Kirzhnits[12] as a quantum correction to the Thomas-Fermi model for atoms,[13,14] this method naturally takes the Thomas-Fermi functional as the zero-order term

$$T_0 \equiv \int t_0(\bm{r})d\bm{r} = T_{TF}[n] = \frac{3}{10}(3\pi^2)^{\frac{2}{3}}\int n^{\frac{5}{3}}(\bm{r})d\bm{r}$$

(2)

and the 1/9th of the von Weizsäcker functional[15] becomes the second-order term

$$T_2 \equiv \int t_2(\mathbf{r})d\mathbf{r} = T_{vW}[n] = \frac{1}{72}\int \frac{|\nabla n(\mathbf{r})|^2}{n(\mathbf{r})}d\mathbf{r}$$

(3)

Further developments of the gradient expansion method by Hodges[16] and Murphy[17] introduced fourth-order and sixth-order corrections respectively:

$$T_4[n] \equiv \int t_4(\mathbf{r})d\mathbf{r} = = \frac{(3\pi^2)^{-\frac{2}{3}}}{540}\int n^{\frac{1}{3}}\left[\left(\frac{\Delta n}{n}\right)^2 - \frac{9}{8}\left(\frac{\Delta n}{n}\right)\left(\frac{\nabla n}{n}\right)^2 + \frac{1}{3}\left(\frac{\nabla n}{n}\right)^4\right]d\mathbf{r}$$

(4)

$$T_6[n] \equiv \int t_6(\mathbf{r})d\mathbf{r} =$$

$$= \frac{(3\pi^2)^{-\frac{4}{3}}}{45360}\int n^{-\frac{1}{3}}\Bigg[13\left(\frac{\nabla\Delta n}{n}\right)^2 + \frac{2575}{144}\left(\frac{\Delta n}{n}\right)^3 + \frac{249}{16}\left(\frac{\nabla n}{n}\right)^2\left(\frac{\Delta^2 n}{n}\right)$$

$$+ \frac{1499}{18}\left(\frac{\nabla n}{n}\right)^2\left(\frac{\Delta n}{n}\right)^2 - \frac{1307}{36}\left(\frac{\nabla n}{n}\right)^2\left(\frac{\nabla n \nabla \Delta n}{n^2}\right) + \frac{343}{18}\left(\frac{\nabla n \nabla \nabla n}{n^2}\right)^2$$

$$+ \frac{8341}{72}\left(\frac{\Delta n}{n}\right)\left(\frac{\nabla n}{n}\right)^4 - \frac{1600495}{2592}\left(\frac{\nabla n}{n}\right)^6\Bigg]d\mathbf{r}$$

(5)

Here we use capital $T$ to define terms in the KEF (integrated over space) and small $t(\mathbf{r})$ for corresponding terms in the KEDF. These corrections had some success. For example, it was shown that the fourth-order gradient expansion $T^{(4)} = T_0 + T_2 + T_4$ gives atomic Hartree-Fock kinetic energy with an error of less than 1% based on Hartree-Fock electron densities,[18,19] which were calculated from analytic Hartree-Fock atomic wave functions by Clementi and Roetti.[20] Although the performance of the $T^{(4)}$ for molecules, calculated with a Gaussian basis, was not so impressive and resulted in errors which were more than two times larger than the errors from the second-order expansion $T^{(2)} = T_{TF} + \frac{1}{9}T_{vW}$,[21] it was shown that the use of a modified expansion with a scaled fourth-order term[9] $T_m^{(4)} = T_{TF} + \frac{1}{9}T_{vW} + \lambda T_4$ improves the accuracy and makes the error less than 1%.[21-23] Apparently, pseudopotentials were not used in those tests, meaning that the results were affected by the strongly corrugated core densities. It is worth of noting, that the use of projector wave-augmented method (PAW)[24] offers an alternative to pseudopotentials way of treating core electron states in the context of orbital-free density functional calculations[25] however this direction of research is not as often used in applications to OF-DFT as the pseudopotential approach.

Numeric convergence issues associated with higher order derivatives and their powers complicate the use of high-order gradient expansion terms. The gradient expansion terms of higher than the second-order can diverge at long range.[26] For example, it was suggested that such behavior makes $T^{(4)}$ unreliable for the description of electron gas models,[27] does not allow for accurate reproduction of non-interacting kinetic energy contribution to the binding energies of molecules,[28] and severely affects the self-consistent minimization procedure of Eq. 1.[26] Attempts to find an approximation that would circumvent this issue have had a degree success in specific cases.[29,30]

The problems with convergence and accuracy are most apparent for calculations of atoms and molecules with an all-electron basis, but they might be not as severe when using pseudopotentials or / and in the case of solid-state periodic calculations, where large low-density regions typical for non-periodic systems are absent, and only the valence density is considered.[31,32] Moreover, it was pointed out that this divergent behavior is global in the sense that each term is an integral over the entire electron distribution, while the respective kinetic energy density can be in general finite except at specific points in space, which can be dealt with separately.[27] Indeed, the consideration of 4th order terms, which were however adapted to satisfy scaling relations, exact conditions or slowly and rapidly varying density limits, has already been shown to substantially improve performance.[33,34]

Overall, one may expect that $T^{(4)}$ can serve as a basis for a more reliable approximation, which nevertheless requires a tuning procedure to find an appropriate set of coefficients for the terms. Further, the sixth order expansion, Eq. 5, was never thoroughly tested neither for molecules nor for solids. The improvements that could be achieved using $T^{(6)}$ with tuned coefficients are also worth considering but are not the subject of the present work.

The aim of the present work is to explore the extent of use of the fourth-order gradient expansions, as published, with tuned coefficients, and when using the terms of the expansion as density dependent variables in a non-linear fit. We consider both solid-state systems and molecules. We do so in the context of PP based calculations, which dominate materials modeling, otherwise the rapidly changing core electron density is expected to limit severely the performance of any KEF approximation. The resulting tuned KED expansions are tested on their ability to reproduce kinetic energy densities as well as potential energy curves of selected molecules and solids.

Albeit the condition of positivity of the Pauli and other properties of kinetic energy desnity,[37,38] can be enforced with the aim to improve non-empirical semi-local functionals,[34,37-40] we do not impose them in our investigation. As has been demonstrated by Levy and Ou-Yang,[37] the condition of positivity of the Pauli term is naturally satisfied with the decomposition of KEDF solely into Thomas-Fermi and full von Weizsäcker terms, but it holds no longer for the decompositions using higher order terms or using a scaled amount of the von Weizsäcker term. Also, imposing the coordinate scaling condition does not necessarily improve machine learned KEDF, as has recently been shown for simplest real system – $H_2$ molecule.[41]

We use an ML approach, specifically neural networks (NN).[42] ML has recommended itself as a useful approach for atomic and molecular simulations by providing extremely accurate fits of a number of parameters to reference property data.[43-46] In the field of orbital-free DFT, ML has already become an actively used tool - the approximation of KED in a one dimension as a function of the electron density by means of kernel ridge regression,[47-49] the approximation for non-local correction to KE by means of convolutional neural networks,[50] direct learning of the potential,[51] and the attempt to construct semi-local KEDF based on up to third-order gradients of the electron density with multi-layer NN's[52] to name but a few recent researches. Specifically, the NN approach has the advantage over some other ML techniques[43] in that it is relatively easy to apply to large data sets (millions of data) used in this work.

At first, we will concentrate on training of NN's for each compound separately to underline some important features concerning pseudopotentials, technical aspects of data distribution and provide reference accuracies. We then explore NN learning simultaneously the KED of multiple components, which is necessary to arrive at a transferable KED functional.

**Methods**

DFT calculations were performed in Abinit 7.10.5[53,54] using the Perdew-Burke-Ernzerhof (PBE) functional.[55] Molecules were placed inside large periodic simulation cells with periodic boundary conditions. For bulk compounds, the unit cells were used. Fermi-Dirac smearing was used in the case of Al, Mg and Li, with the corresponding smearing of 0.01 Ha. The resulting lattice parameters together with other calculations details are listed in Table 1. The plane-wave cutoffs were chosen based on convergence tests. The used *k*-point sampling grids have provided converged results. The force tolerance was equal to 5×10$^{-6}$ Hartree/Bohr. The energy tolerance for self-consistent minimization procedure was equal to 10$^{-10}$ Hartree. Goedeker-Teter-Hutter (GTH) pseudopotentials[57,58] were used for Al, Mg, and Li (the case of a large core for former two), while for O, H, C, Si and for Mg and Li with small cores Fritz-Haber-Institute (FHI) pseudo-potentials[59,60] were taken.

The fitting was performed in Matlab Neural Network Toolbox.[61] Input vectors for fitting comprised five elements $t_0, t_2, t_4^{(1)}, t_4^{(2)}, t_4^{(3)}$, all of them were terms entering $T^{(4)}$:

$$t_0[n(\boldsymbol{r})] = n^{\frac{5}{3}}(\boldsymbol{r})$$

$$t_2[n(\boldsymbol{r})] = \frac{1}{8}\frac{|\nabla n^2|}{n} - \frac{1}{4}\Delta n(\boldsymbol{r})$$

$$t_4^{(1)}[n(\boldsymbol{r})] = n^{\frac{1}{3}}(\boldsymbol{r})\left(\frac{\Delta n}{n}\right)^2$$

$$t_4^{(2)}[n(\boldsymbol{r})] = n^{\frac{1}{3}}(\boldsymbol{r})\left(\frac{\Delta n}{n}\right)\left(\frac{\nabla n}{n}\right)^2$$

$$t_4^{(3)}[n(\boldsymbol{r})] = n^{\frac{1}{3}}(\boldsymbol{r})\left(\frac{\nabla n}{n}\right)^4$$

(6)

Table 1. Calculation details.

| Compound | Optimized lattice constants, a.u. | Experimental lattice constants,[56] a.u. | Energy cutoff, Ha | $k$-point sampling | Sampling resolution of real-spaced grid, a.u. | Number of test and fitting points |
|---|---|---|---|---|---|---|
| Aluminum | 7.542 | 7.652 | 16.0 | 8×8×8 | 0.0250 | 9.2×10$^6$ |
| Magnesium | a=6.080[a] | a=6.065 | 20.0 | 8×8×4 | 0.0334 | 3.4×10$^6$ |
| | c=9.872[a] | b=9.847 | | | | |
| | a=6.127[b] | | | | | |
| | c=9.947[b] | | | | | |
| Lithium | 6.463[c] | 6.633 | 16.0 | 8×8×8 | 0.0334 | 2.7×10$^6$ |
| | 6.686[d] | | | | | |
| Silicon | 10.327 | 10.263 | 24.0 | 8×8×8 | 0.0251 | 9.4×10$^6$ |
| H$_2$O | 15.118 | - | 20.0 | Γ point | 0.0501 | 9.15×10$^6$ |
| C$_6$H$_6$ | 18.897 | - | 20.0 | Γ point | 0.0665 | 7.6×10$^6$ |

[a] 2 valence electrons; [b] 10 valence electrons; [c] 1 valence electron; [d] 3 valence electrons.

In neural network configurations the transfer function between last hidden and output layer is linear. In this case NN equation is

$$KED(t) = d + \sum_{n=1}^{N_{neurons}} c_n \sigma(\mathbf{w_n} \cdot \mathbf{t} + b_n)$$

(7)

Where $\mathbf{t}$ is the vector of the five input variables $t_i$ defined above (Eq. 6) and $N_{neurons}$ is the number of neurons in the hidden layer. When a single *linear* neuron is used in the hidden layer ($\sigma(x) = x$), NN training is reduced to linear regression (linear fit of expansion coefficients), this is designated as "fitted" below. Non-linear NN fitting is designated as "$N_{neurons}$ NN" and the neuron activation function in this case is sigmoidal ($\sigma(x) = \frac{e^x - e^{-x}}{e^x + e^{-x}}$). For NN training, the values at the grid points were mapped to the interval [-1.0, 1.0]. The distributions of the kinetic energy density values for the studied systems are shown in Figure 1.

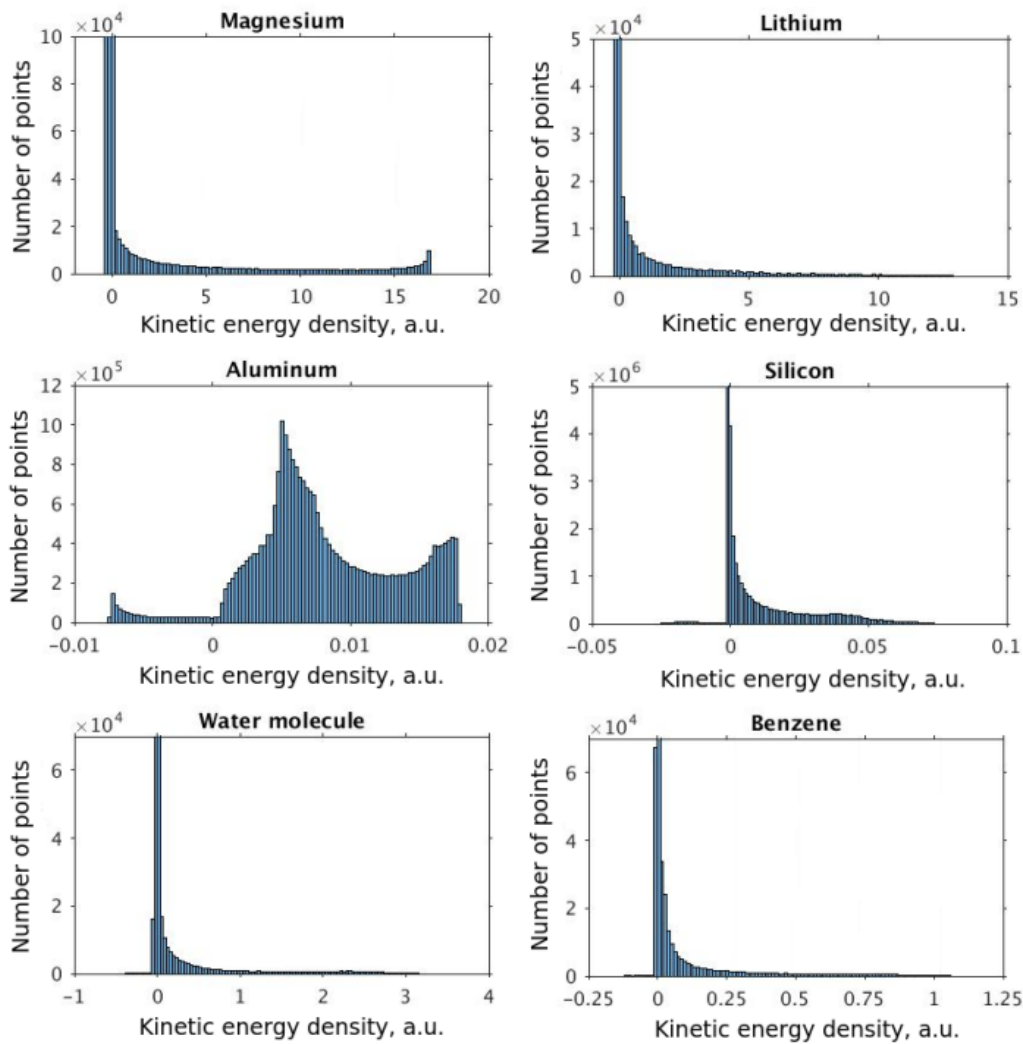

Figure 1. Distribution of kinetic energy densities.

It is seen that the distributions are strongly non-uniform, which makes the data difficult to fit. Therefore, in order to ensure more appropriate coverage for each part of the distributions, the fit and test data were selected in the following way: four intervals were defined for scaled values of KED: [-1.0, -0.5], [-0.5, 0.0], [0.0, 0.5], and [0.5, 1.0]. The fraction of cube points drawn from each KED interval for use in fit and test sets was the higher the smaller the number of values in the corresponding part of the histogram of Figure 1. If the number of data in a given interval was smaller than 1/5000th of the total number of points, all of them were drawn. This is a similar approach to histogram equalization.[63] However, when systems with more complex than near-uniform (Li, Mg, Al) electron density profile are considered, even the above described approximate histogram equalization scheme is not sufficient. For example, in our investigation this was the case for Si, where the numerical instabilities arose in NN-fitted KES near the nucleus due to the relatively low number of sampling points, even though there was no overfitting as the total number of data was overwhelmingly large compared to the number of NN parameters (see Figure 1, silicon, the region with negative values of KED). Further equalization then was performed by producing additional cube files just around Si atoms with an approximately 2.5 times higher resolution. Thus, having in such a way two sets of cube files (one for the unit cell, another one for the near-atomic region), one half of test and fitting points was drawn from the first set, another half was drawn from the second set. The NN fitting was done to minimize the root mean square error (RMSE) of the kinetic energy density.

OFDFT calculations were carried out in PROFESS 3.0[64] using PBE exchange-correlation functional with a plane wave cutoff of 3600 eV, which provided converged results.

## Results

### The performance of the formal gradient density expansion

The crystal structures of Li (bcc), Al (fcc), Mg (hcp), Si (cubic diamond) as well as the molecular structures of $H_2O$ and $C_6H_6$ are shown in Figure 2. The lines along which the representative plots of the kinetic energy density computed with different approximations have the following directions: [100] for Al, [111] for Si, and along the line between atoms at (0, 0, 0) and (2/3, 1/3, 1/2) Wyckoff positions for Mg. For molecules, the lines are along interatomic bonds. The intervals over which KED are plotted in Figure 3 - Figure 10 are shown as arrows in Figure 1. We also examined 2D slices of the KED. Quality of KED approximation with different 4th order expansions are also shown in Figure 3 - Figure 10 on planes (passing through atoms) {1122} for Mg, {101} for Li, {101} for Al, and {111} for Si, as well as molecular planes.

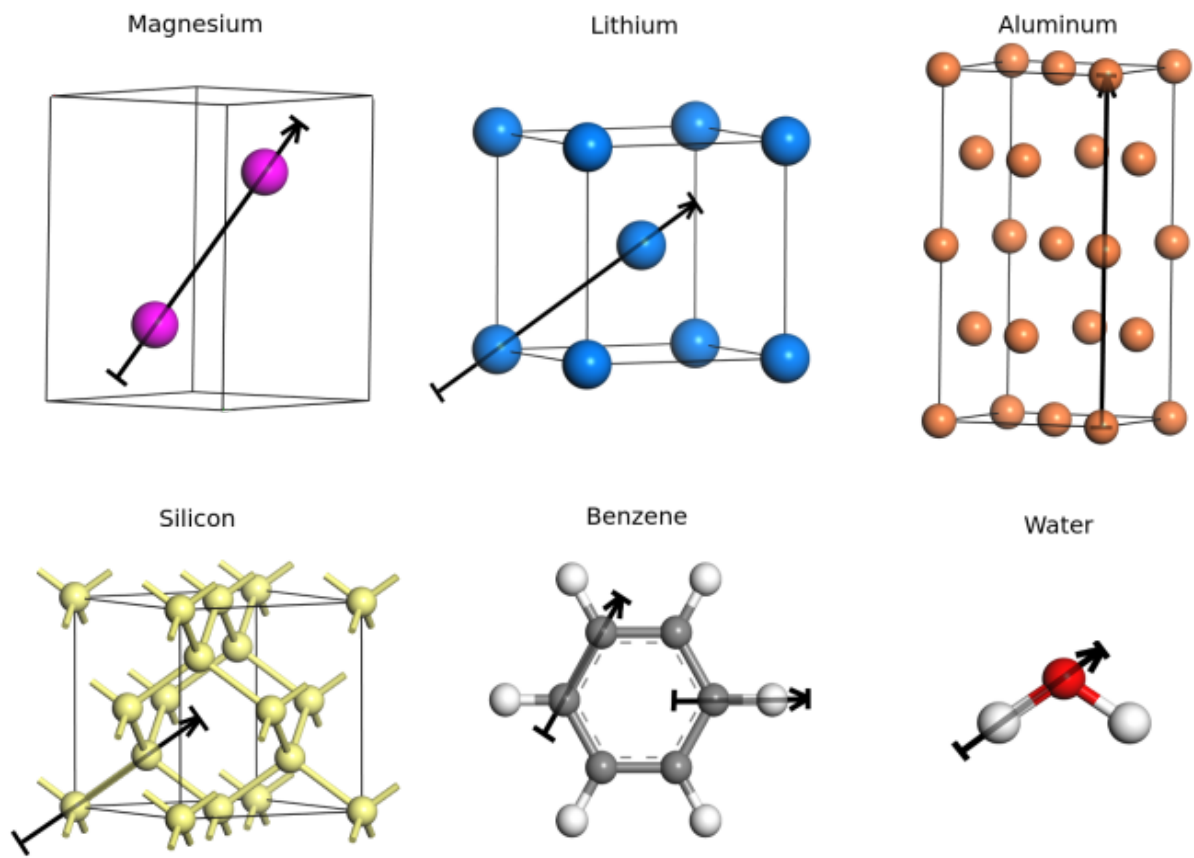

Figure 2. The crystal structures of Li, Al, Mg, Si as well as the molecular structures of $H_2O$ and $C_6H_6$. The arrows indicate the directions and intervals along which the kinetic energy density is plotted in Figure 3 - Figure 10. Unit cells are shown for Li and Mg, and 2 unit cells for Al.

The 1D plots of kinetic energy densities in selected directions are shown in Figure 3 - Figure 10 together with 2D plots of errors vs KS KED with different models. The difference between the $2^{nd}$- and the $4^{th}$-order expansions is noticeable mostly at the regions close to nuclei, where the corresponding valence electron densities tend to their minimum. The tendency is pseudopotential-dependent and may give rise to numeric instabilities (as is clearly seen in the case of Li and Mg) discussed in the next subsection. From the curves of kinetic energy densities, it is seen that the inclusion of $4^{th}$-order terms $t_4^{(1-3)}$ affects mostly the near-nuclei regions, significantly increasing the values of KED there. It is often the case when the local minimum of the $t_2$ formal expansion turns in this way to a local maximum of the formal $t_4$ expansion. We use "formal" for the expansions of Eqs. 4 and 6, as opposed to the tuned expansions presented below. While in some cases this behavior is in accordance with the general shape of KS kinetic energy density curves, there are the cases where it is not – for example, water molecule and benzene, Figure 9 and Figure 10, respectively. In the regions of space far from nuclei, the formal $t_2$ and $t_4$ expansions tend to match. Table 2 clearly shows that for the light metals considered here, the 4$^{th}$ order expansion does improve the KED, for the covalently bonded bulk Si, there is no improvement in the total RMSE, while in the molecular systems, there is significant worsening of the quality of KED upon inclusion of the 4$^{th}$ order terms. In both Si and the molecules, $t_4$ fails in regions where the data is scarce (see Figure 1). It is in principle possible that $T^{(4)}$ would perform better under different data distributions.

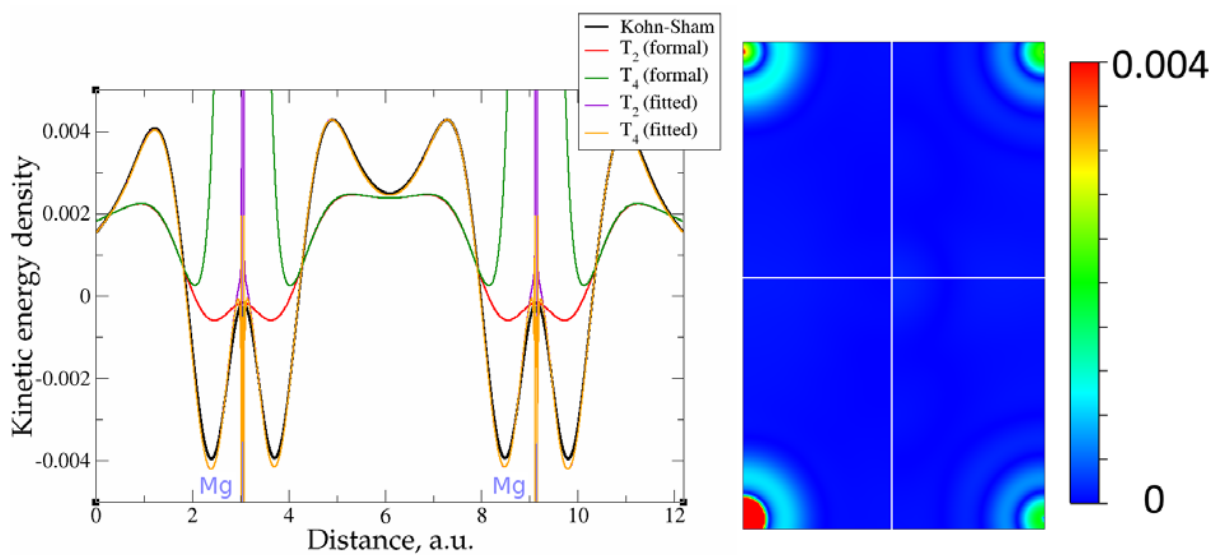

Figure 3. Kinetic energy density for magnesium (2 valence electrons), computed with Kohn-Sham DFT and with different gradient approximations for the KEDF (applied to the KS DFT density). On the left – the line along atoms at (0, 0, 0) and (2/3, 1/3, 1/2) Wyckoff positions; on the right – the projection of plane {1 1 2 2} with absolute differences between KEDs of different gradient approximations and Kohn-Sham KED: upper left quadrant – formal $T^{(2)}$, upper right quadrant – fitted $T^{(2)}$, bottom left quadrant – formal $T^{(4)}$, bottom right quadrant – fitted $T^{(4)}$.

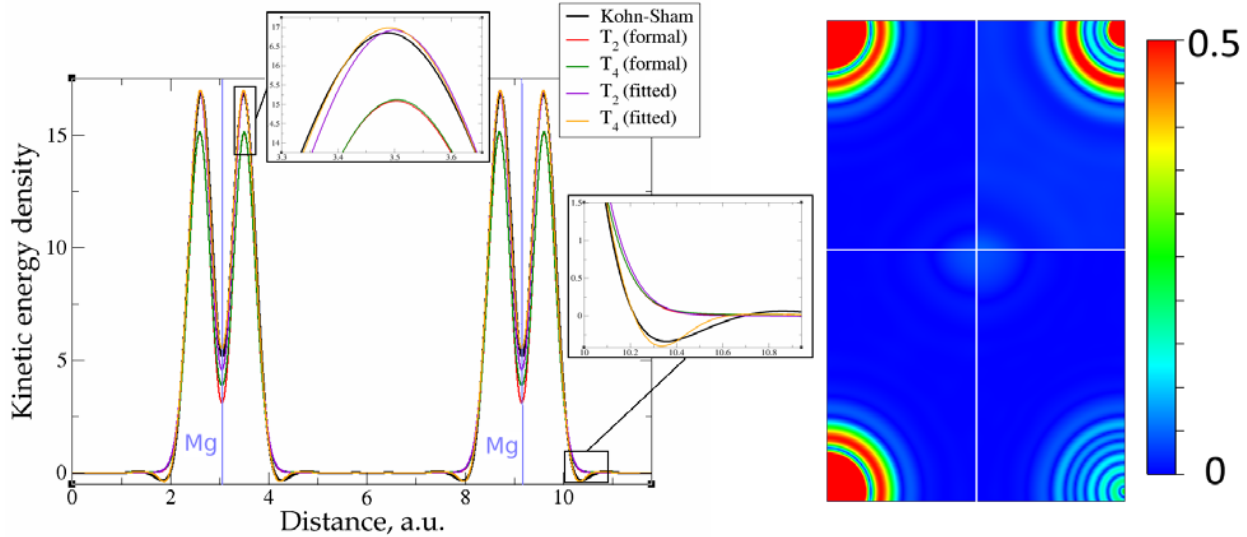

Figure 4. Kinetic energy density for magnesium (10 valence electrons), computed with Kohn-Sham DFT and with different gradient approximations for the KEDF (applied to the KS DFT density). On the left – the line along atoms at (0, 0, 0) and (2/3, 1/3, 1/2) Wyckoff positions; on the right – the projection of plane {1 1 2 2} with absolute differences between KEDs of different gradient approximations and Kohn-Sham KED: upper left quadrant – formal $T^{(2)}$, upper right quadrant – fitted $T^{(2)}$, bottom left quadrant – formal $T^{(4)}$, bottom right quadrant – fitted $T^{(4)}$. For the better visibility, the scale for 2D plot is different form the one used for Figure 3.

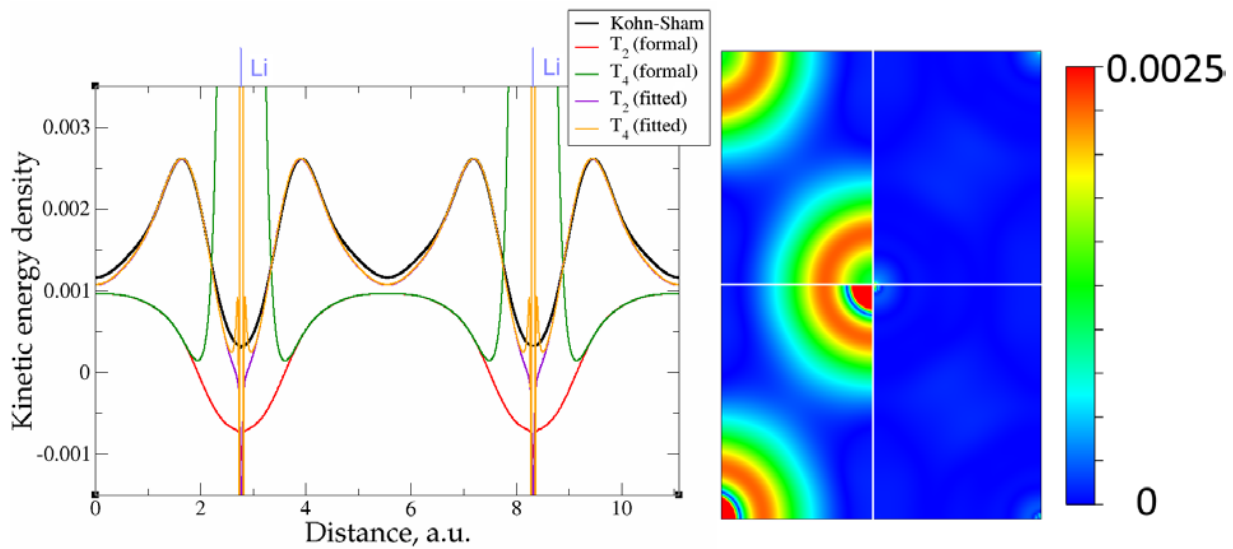

Figure 5. Kinetic energy density for lithium (1 valence electron), computed with Kohn-Sham DFT and with different gradient approximations for the KEDF (applied to the KS DFT density). On the left – the line along direction {1 1 1}; on the right – the projection of plane {1 0 1} with absolute differences between KEDs of different gradient approximations and Kohn-Sham KED: upper left quadrant – formal $T^{(2)}$, upper right quadrant – fitted $T^{(2)}$, bottom left quadrant – formal $T^{(4)}$, bottom right quadrant – fitted $T^{(4)}$.

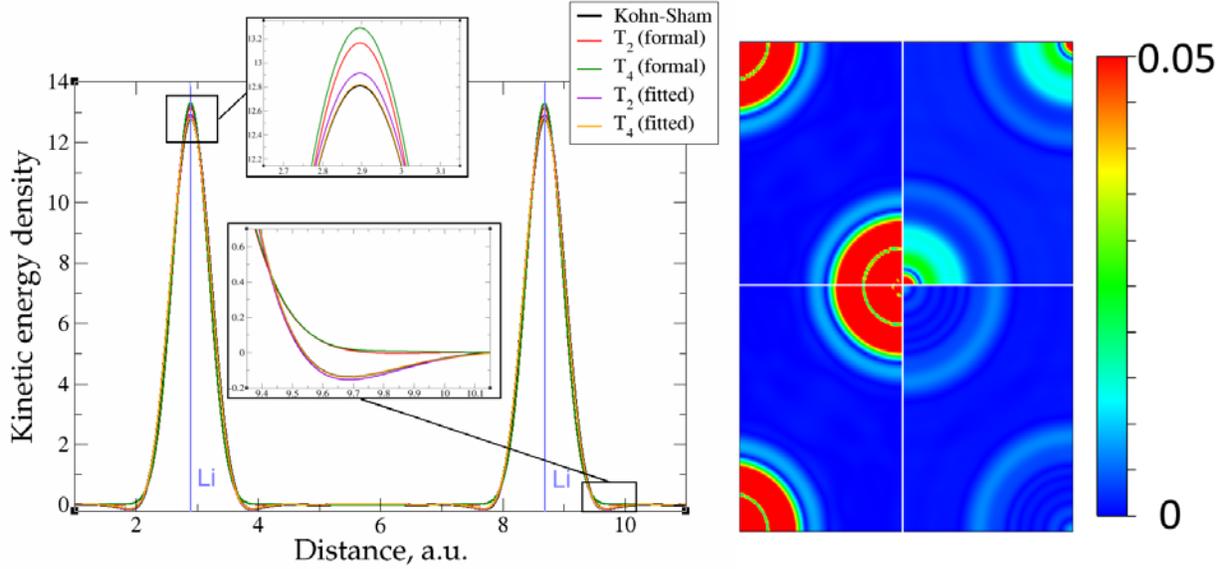

Figure 6. Kinetic energy density for lithium (3 valence electrons), computed with Kohn-Sham DFT and with different gradient approximations for the KEDF (applied to the KS DFT density). On the left – the line along direction {1 1 1}; on the right – the projection of plane {1 0 1} with absolute differences between KEDs of different gradient approximations and Kohn-Sham KED: upper left quadrant – formal $T^{(2)}$, upper right quadrant – fitted $T^{(2)}$, bottom left quadrant – formal $T^{(4)}$, bottom right quadrant – fitted $T^{(4)}$. For better visibility the scale for 2D plot is different form the one used for Figure 5.

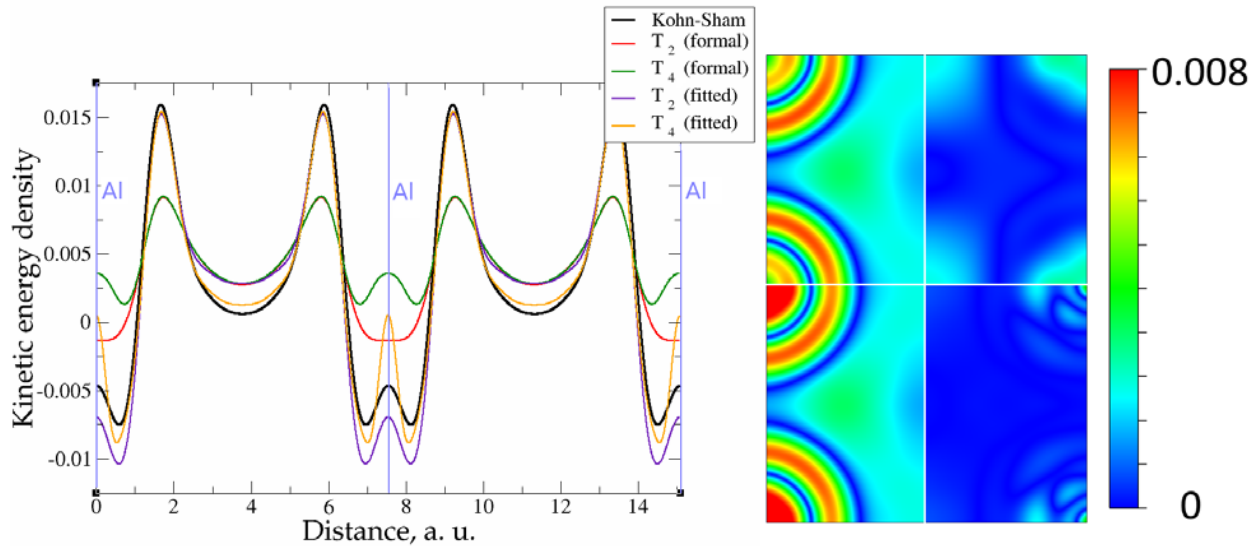

Figure 7. Kinetic energy density for aluminum (3 valence electrons), computed with Kohn-Sham DFT and with different gradient approximations for the KEDF (applied to the KS DFT density). On the left – the line along direction {1 0 0}; on the right – the projection of plane {1 0 1} with absolute differences between KEDs of different gradient approximations and Kohn-Sham KED: upper left quadrant – formal $T^{(2)}$, upper right quadrant – fitted $T^{(2)}$, bottom left quadrant – formal $T^{(4)}$, bottom right quadrant – fitted $T^{(4)}$.

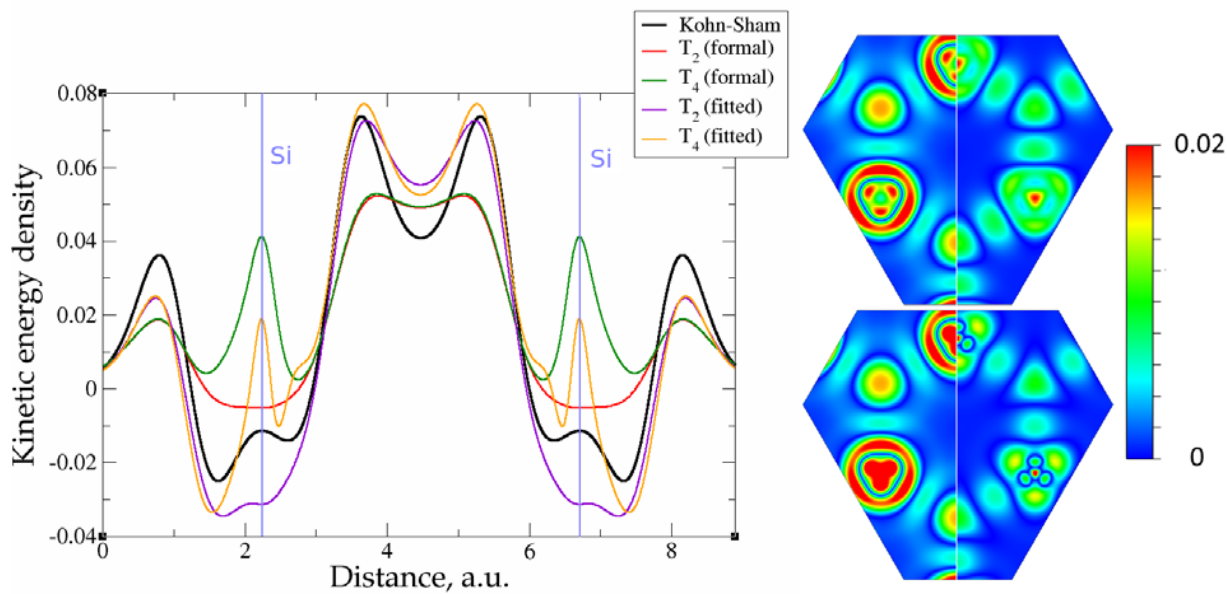

Figure 8. Kinetic energy density for silicon, computed with Kohn-Sham DFT and with different gradient approximations for the KEDF (applied to the KS DFT density). On the left – the line along direction {1 1 1}; on the right – the projection of plane {1 1 1} with absolute differences between KEDs of different gradient approximations and Kohn-Sham KED: upper left quadrant – formal $T^{(2)}$, upper right quadrant – fitted $T^{(2)}$, bottom left quadrant – formal $T^{(4)}$, bottom right quadrant – fitted $T^{(4)}$.

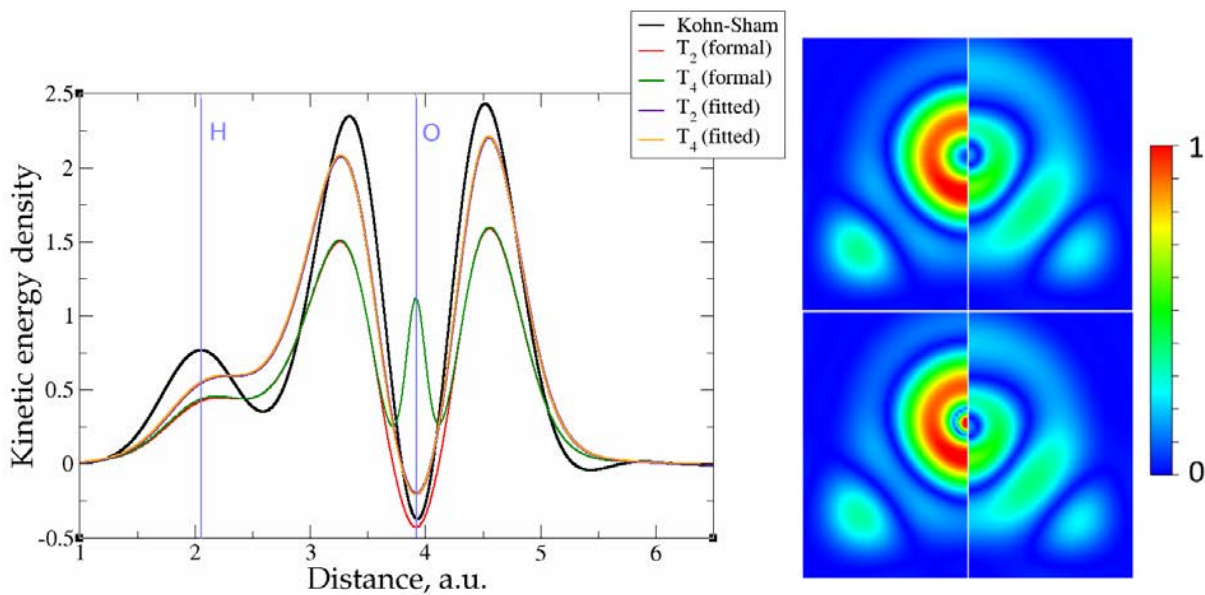

Figure 9. Kinetic energy density for water molecule, computed with Kohn-Sham DFT and with different gradient approximations for the KEDF (applied to the KS DFT density). On the left – the line along O-H bond; on the right – the projection of molecular plane with absolute differences between KEDs of different gradient approximations and Kohn-Sham KED: upper left quadrant – formal $T^{(2)}$, upper right quadrant – fitted $T^{(2)}$, bottom left quadrant – formal $T^{(4)}$, bottom right quadrant – fitted $T^{(4)}$.

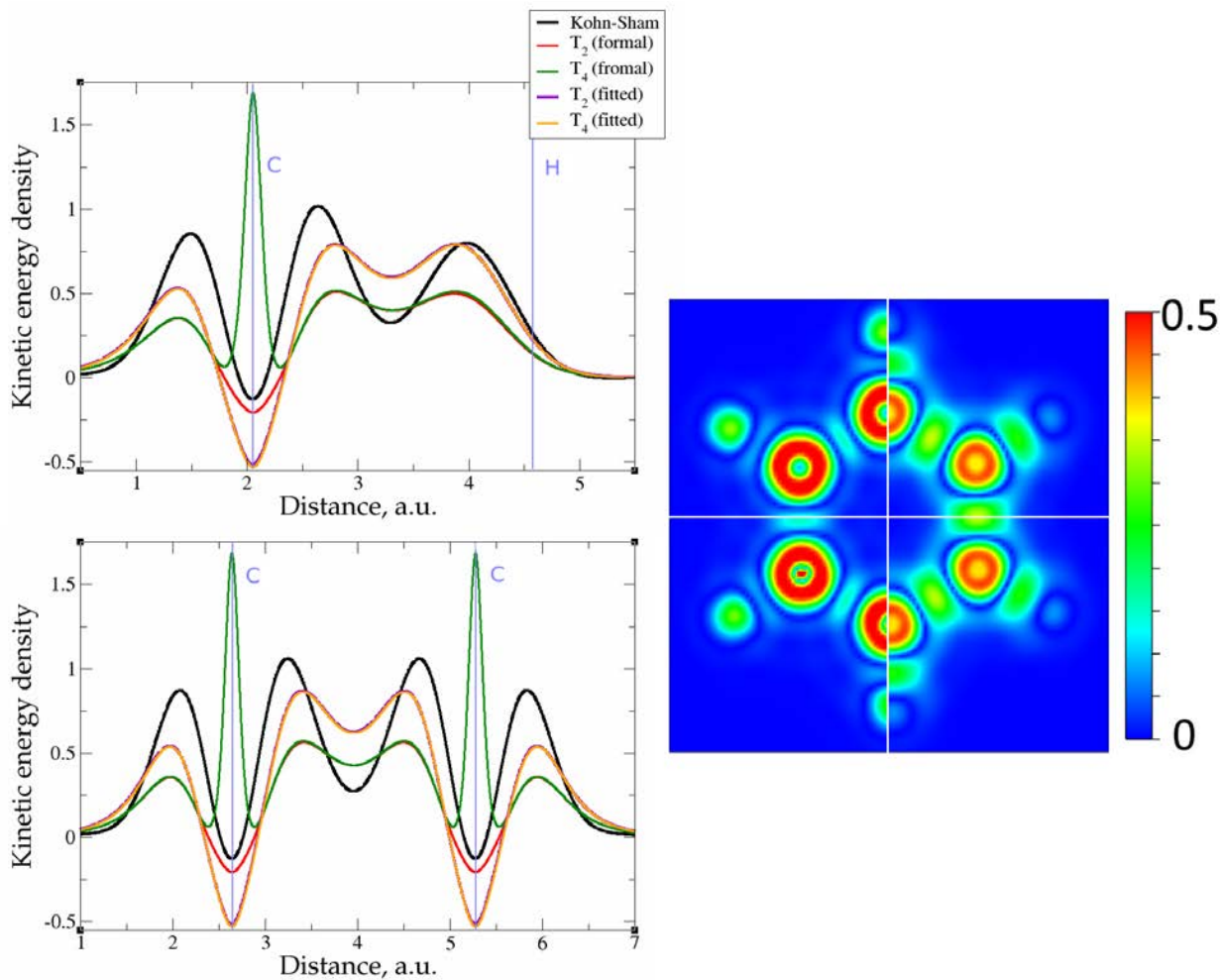

Figure 10. Kinetic energy density for benzene, computed with Kohn-Sham DFT and with different gradient approximations for the KEDF (applied to the KS DFT density). Form the left – the lines along C-H and C-C bonds; on the right – the projection of molecular plane with absolute differences between KEDs of different gradient approximations and Kohn-Sham KED: upper left quadrant – formal $T^{(2)}$, upper right quadrant – fitted $T^{(2)}$, bottom left quadrant – formal $T^{(4)}$, bottom right quadrant – fitted $T^{(4)}$.

Overall, one cannot conclude that the formal $T^{(4)}$ expansion systematically improves over the formal $T^{(2)}$ expansion. The failure of the formal density gradient expansions to reproduce quantitatively the KS kinetic energy density was expected. It was also expected that the expansions would perform much better for light metals with a more uniform density than for molecules. Below, we explore possibilities of using $T^{(4)}$ as a basis for the further improvements.

## The role of pseudopotentials

Among the examples covered in this work, there are two cases with numerical instability of gradient density expansions in close proximity to the nuclei: Mg with two valence electrons (Figure 3), and Li with 1 valence electron (Figure 5). The reason is small, almost near-zero, values of valence electron densities at the nuclei that affect the terms with electron densities of different powers in denominators. What is important is not just the low density but how it tends to the minimum, which is reflected in the respective electron density derivatives. Indeed, considering, for example, the water molecule (Figure 9), one sees that there are no instabilities in the regions far from the nuclei, where the electron density also reaches near-zero values but in a more gradual way.

The shape of the valence electron density is governed by the respective pseudopotential. For modern non-local pseudopotentials, it is often the case that the *s*-electron pseudopotential channel is characterized by a relatively high positive value (on the order of 1 a.u. and higher) at the origin (see, for example, Figure 11 for a GTH pseudopotential for Mg with 2 valence electrons). This channel also dominates local PPs for elements with s-only valence shell.[65,66] The density then decays exponentially fast and steeply towards the nucleus. This forces the expressions with the electron density of different powers in the denominator and powers of derivatives in the numerator to become numerically unstable around the nucleus. Sometimes, this affects even the numerically stable representation of $T_2$, without the electron density in the denominator,[64]

$$t'_2 = -\frac{1}{2}\int \sqrt{n(\boldsymbol{r})}\Delta\sqrt{n(\boldsymbol{r})}d\boldsymbol{r}$$

(7)

For example, with a GTH pseudopotential for lithium, this expression also results in near-nuclei numerical instabilities, this time clearly caused by the second-order derivative.

A similar issue, together with what was called by authors an "uncertainty" of pseudopotentials, was reported by Karasiev and Trickey[65] in relation to GGA[68] and mcGGA[69] functionals with the square of reduced density as one of the components. For the mcGGA functional this was attributed, at least partially, to the imposition of positivity constrain.[70] Later the way to overcome this problem in the context of GGA-type KEDF's was proposed.[71]

However, the issue we face looks to be purely technical; it is related to elements with only *s*-electrons in valence shell and is not dependent on any constrains. Indeed, the presence of the contribution from *p* valence electrons, for which the corresponding valence *p*-electron pseudopotential channel is normally negative at the origin, is able to make the total valence electron density approach its minimum at the nucleus in a more gradual way. For example, no numerical instabilities appeared in the case of aluminum with three valence electrons (see Figure 7), whose valence shell differs from that Mg only by one *p* electron.

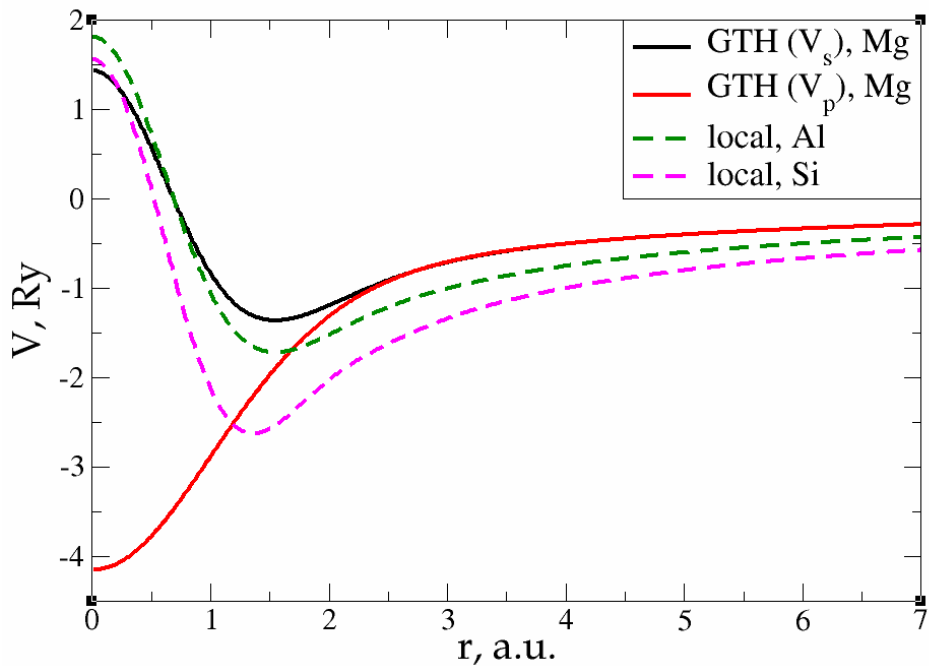

Figure 11. *s*- and *p*-channels of a GTH pseudopotential for Mg with 2 valence electrons as well as local pseudopotentials for Al and Si.[67]

This numerical issue is resolved by increasing the number of explicitly considered electrons (so call "semi-core" pseudopotential), for example, by using 10 electrons for Mg or 3 electrons for Li (see Figure 4 and Figure 6). In addition, there is no formal reason that would prohibit producing pseudopotentials that would force the *s*-electron valence densities to approach a minimum at the nucleus more gradually, thus enabling the consideration of only the valence *s*-states for alkali and alkaline earth metals. This is something that pseudopotential development for OF-DFT might consider in the future.

In summary, with the proper choice of pseudopotential, in the framework of solid-state periodic calculations, the $T^{(4)}$ expansion terms remain continuous everywhere and can serve as a base for further search of an accurate approximation to kinetic energy density.

## NN training on KEDs of individual compounds

For individual fits we used a single hidden layer neural network, which is a universal approximator.[43,72,73] The number of training and test points was chosen to be approximately equal to one-third of the total number of sampling points each.

## Coefficients fitting

The utility of different-order gradient expansions can be explored by fitting of the corresponding term coefficients for each of the studied systems. That might answer the question if it is possible at all to find a linear set of coefficients that allows for a close approximation to the KS kinetic energy density for a broad range of compounds. Fitting of the coefficients results in much improvement over the formal $T^{(2)}$ and $T^{(4)}$ gradient density expansions, which can be detected even visually, as can be seen in Figure 3 - Figure 10. The most successful cases are elemental metals – Li, Mg, and Al. For them fitted $T^{(4)}$ in general is able to produce a better fit to $T_{KS}$ than fitted $T^{(2)}$. That is most noticeable in the interatomic region in Al (Figure 7) and near-nuclear regions in Mg and Li (Figure 4 and Figure 6, respectively). Only the fitted $t_4$ was able to reproduce a local minimum of $t_{KS}$ for Mg (10 valence electrons) and Li (3 valence electrons) in regions of both large and small values of $t_{KS}$ (see the inserts of Figure 4 and Figure 6, respectively). Although there are some regions where $t_4$ seemingly performs worse than $t_2$, in other regions, significant improvements are observed. It might be that there is not enough freedom in linear fitting to achieve improvements with $T^{(4)}$ terms in all regions of space. The performance may also be affected by data distribution. Nevertheless, the overall fitting performance of $T^{(4)}$ is (necessarily) better, as can be seen from the RMSE on the entire test set (Table 2). However, coefficients fitting procedure was not able to fix the above-discussed numerical instabilities associated with pseudopotentials.

As expected, for molecules and Si, the fit performs worse than for elemental metals, which questions the suitability of gradient expansions to reproduce both near-nuclei and covalently bonded regions. Moreover, for these compounds $T^{(4)}$ does not improve the KED much over $T^{(2)}$, contrary to the elemental metals.

Table 2. Root mean square error of kinetic energy density (Ha/Bohr$^3$) over test data for 2$^{nd}$- and 4$^{th}$-order expansions.

| Expansion | Al | Mg$^a$ | Li$^b$ | Si | H$_2$O | C$_6$H$_6$ |
|---|---|---|---|---|---|---|
| $T^{(2)}$ | 2.07×10$^{-2}$ | 3.03 | 1.01 | 1.65×10$^{-2}$ | 2.05×10$^{-2}$ | 2.06×10$^{-2}$ |
| $T^{(4)}$ | 1.17×10$^{-2}$ | 1.78 | 0.95 | 1.67×10$^{-2}$ | 1.65×10$^{-1}$ | 8.53×10$^{-2}$ |
| Fitted $T^{(2)}$ | 8.19×10$^{-4}$ | 1.01×10$^{-1}$ | 4.78×10$^{-3}$ | 3.96×10$^{-3}$ | 1.60×10$^{-2}$ | 1.94×10$^{-2}$ |
| Fitted $T^{(4)}$ | 3.33×10$^{-4}$ | 2.43×10$^{-2}$ | 2.26×10$^{-3}$ | 3.57×10$^{-3}$ | 1.60×10$^{-2}$ | 1.94×10$^{-2}$ |

$^a$ 10 valence electrons; $^b$ 3 valence electrons

It is worth noting that in the case of elemental metals (Li, Mg, Al) the systematic attempt of the fitting procedure to make $t_0$ contribution negative (in the range from -0.05 to -0.20, here and below this refers to scaled KED values), when only the valence electrons are considered (1, 2 and 3 electrons for Li, Mg, and Al, respectively). The $t_0$ contribution becomes positive when 1$s$ (Li) or 2$s$2$p$ (Mg) states are included in the pseudopotential. However, in that case, the value of the $t_0$ coefficient is still significantly smaller than in the formal expansion (of around 0.05 for Li and 0.96 for Mg). In contrast, the $t_2$ contribution becomes in general larger – from 1/9 in the formal expansions to 0.95-1.27 in the case of Al and Li and even 4.28 in for Mg with 2 valence electrons. This tendency is also observed for other systems – the $t_2$ contribution is always larger than in the formal expansion, and the $t_0$ contribution is smaller and always positive. This is in line with suggestion that Thomas-Femi term includes a part of KED incorrectly, and that the performance of gradient density expansions might be improved by subtracting this part.[74,75] It is also has been shown that the usage of a larger fraction of $t_2$ in 2nd order gradient expansion gives closer total energies to Hartree-Fock references in the case of noble gas atoms (Ne, Ar, Kr, Xe)[76] and closer binding energies in the case di-atomic molecules (H$_2$, N$_2$, O$_2$, HF and CO).[77] Overall, the test on atoms up to 3rd row inclusive suggests that the higher the fraction of $t_2$, the lower the fraction of $t_0$ that should be used in order to reproduce correctly the total energy; however, this dependence is not linear.[78] It is hard to discern a common pattern in behavior of the coefficient for $t_4^{(1-3)}$ terms. The terms might have either positive or negative sign depending on compound. However, what is common is that their contributions to the KED are one-two orders of magnitude larger than in the formal expansion.

In spite of the apparent improvements offered by the fitting procedure, the results for molecules and systems with localized electrons (Si) remain unsatisfactory. On the other hand, these tests show that systems with delocalized electrons (metals) and a simpler topology of $t_{KS}$ can be quite well reproduced. The next step for the improvement of gradient density expansions is to go beyond coefficient tuning, and this is considered next.

## Non-linear fitting with neural networks using terms of the 4th order expansion as density-dependent variables

In the previous section, it was shown that the 4th order expansion either results in a better match to $T_{KS}$ or does not perform worse than the fitted 2nd order expansion. For covalently bound systems, however, fitted $T^{(4)}$ did not result in improvement in KED over fitted $T^{(2)}$. We therefore fitted KEDFs by training neural networks on inputs that were terms in the $T^{(4)}$ expansion. Table 3 lists RMSE errors of kinetic energy density integrated over test data for NN fits using different numbers of hidden neurons. There is a significant improvement in the KE over linearly tuned coefficients. This improvement is obvious also for the covalently bound bulk Si and water and benzene molecules. Most significant improvements are for Al and Mg where the RMSE with 12 neurons becomes smaller by an order of magnitude than the RMSE in the case of linear fitting. For other compounds, the improvements are also significant, leading to 4-5 times better test set RMSEs.

As expected, the fitting precision is systematically increased with the increase of the number of neurons $N_n$ for small $N_n$. For covalently bound systems, the test error can increase for $N_n$ larger than 10. This likely has to do with the distribution of data. While the total dataset contains millions of entries (i.e. a much larger number than the number of NN parameters thus ensuring that NN coefficients are not overfitted), there are few data in specific ranges of the KED (Figure 1). The effect of non-linear fitting can be visually observed in Figure 12 - Figure 17. For metals, NN-based KED can be made follow $t_{KS}$ very accurately. For water and benzene molecules, NN-based KED appears to follow $t_{KS}$ very accurately in regions where there are sufficient training data; it fails badly in regions of few data. Data distribution therefore appears to be a major issue in covalently bound systems. A way to improve the fitting was demonstrated in the case of Si (see 'Methods') - the 'poorly' represented regions of KED values can be replenished from another denser point grid that is built just around these regions (atoms or molecules). It should be noted, that simple increase of grid points resolution for the whole simulation cell is not enough, since the 'problematic' regions of KED values will be still 'poorly' represented relatively to other regions, and the corresponding histogram of KED values distribution will remain strongly non-uniform. More advanced techniques to deal with data distribution likely have to be developed for successful application of gradient expansion based KEDF in such systems.

Table 3. Root mean square error of kinetic energy density (Ha/Bohr$^3$) over test data for neural network fits with different numbers of neurons $N_n$ based on the 4$^{th}$ order expansion.

| $N_n$ | Al | Mg[a] | Li[b] | Si | H$_2$O | C$_6$H$_6$ |
|---|---|---|---|---|---|---|
| 2 | 1.79×10$^{-4}$ | 6.43×10$^{-3}$ | 1.23×10$^{-3}$ | 2.73×10$^{-3}$ | 4.57×10$^{-3}$ | 1.50×10$^{-2}$ |
| 4 | 3.86×10$^{-5}$ | 3.04×10$^{-3}$ | 9.24×10$^{-4}$ | 2.16×10$^{-3}$ | 3.55×10$^{-3}$ | 1.27×10$^{-2}$ |
| 6 | 2.07×10$^{-5}$ | 1.55×10$^{-3}$ | 1.04×10$^{-3}$ | 1.47×10$^{-3}$ | 3.42×10$^{-3}$ | 1.22×10$^{-2}$ |
| 8 | 1.34×10$^{-5}$ | 1.41×10$^{-3}$ | 1.00×10$^{-3}$ | 1.62×10$^{-3}$ | 3.62×10$^{-3}$ | 1.15×10$^{-2}$ |
| 10 | 1.41×10$^{-5}$ | 8.38×10$^{-4}$ | 5.78×10$^{-4}$ | 1.35×10$^{-3}$ | 3.06×10$^{-3}$ | 1.19×10$^{-2}$ |
| 12 | 1.27×10$^{-5}$ | 7.26×10$^{-4}$ | 1.01×10$^{-3}$ | 7.07×10$^{-4}$ | 3.39×10$^{-3}$ | 1.19×10$^{-2}$ |

[a] 10 valence electrons; [b] 3 valence electrons

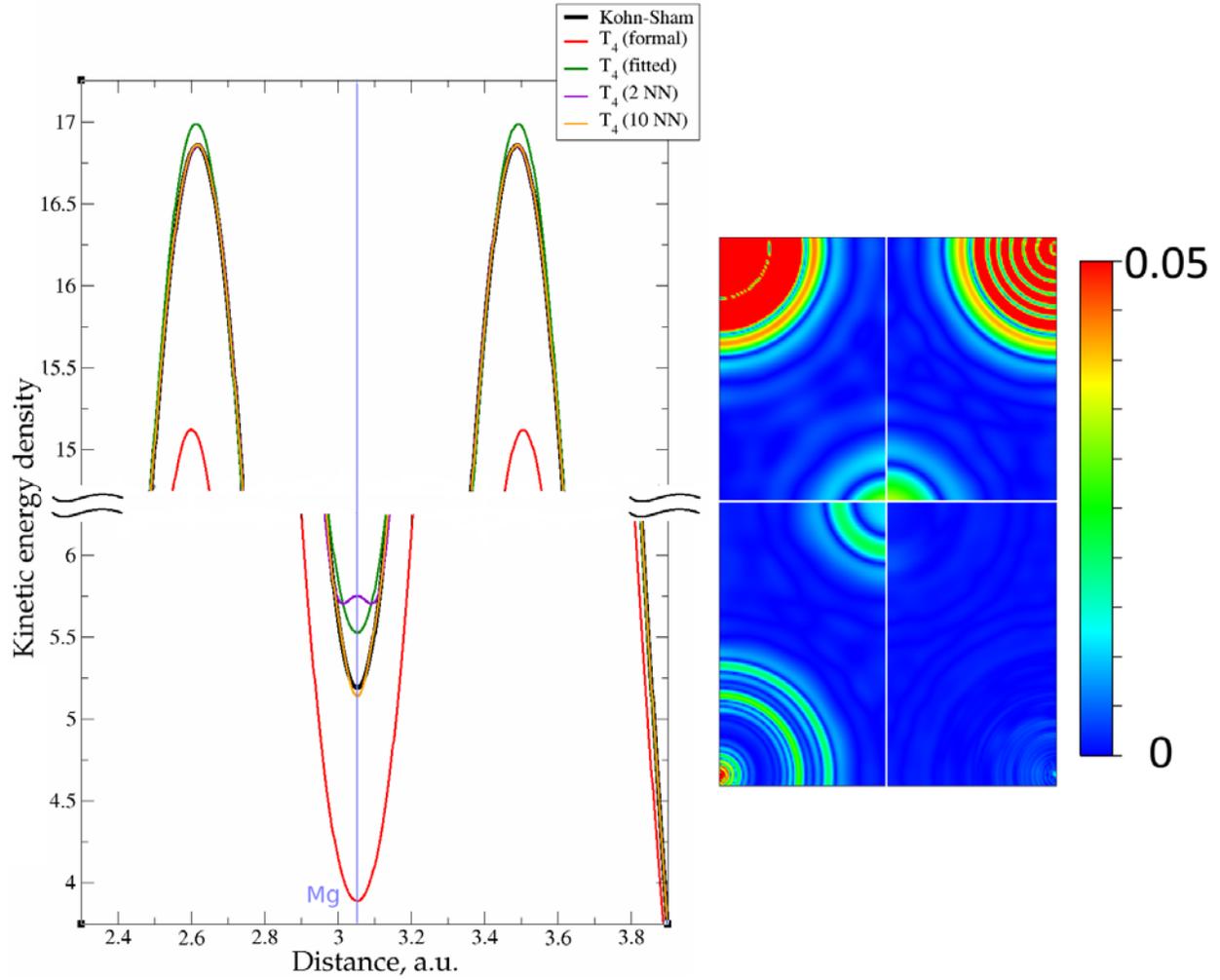

Figure 12. Kinetic energy density from non-linear (NN) fitting for magnesium (10 valence electrons). On the left – the line along atoms at (0, 0, 0) and (2/3, 1/3, 1/2) Wyckoff positions; on the right – the projection of plane {1 1 2 2} with absolute differences between KEDs of different gradient approximations and Kohn-Sham KED: upper left quadrant – formal $T^{(4)}$, upper right quadrant – fitted $T^{(4)}$, bottom left quadrant – 2 neurons $T^{(4)}$, bottom right quadrant – 10 neurons $T^{(4)}$.

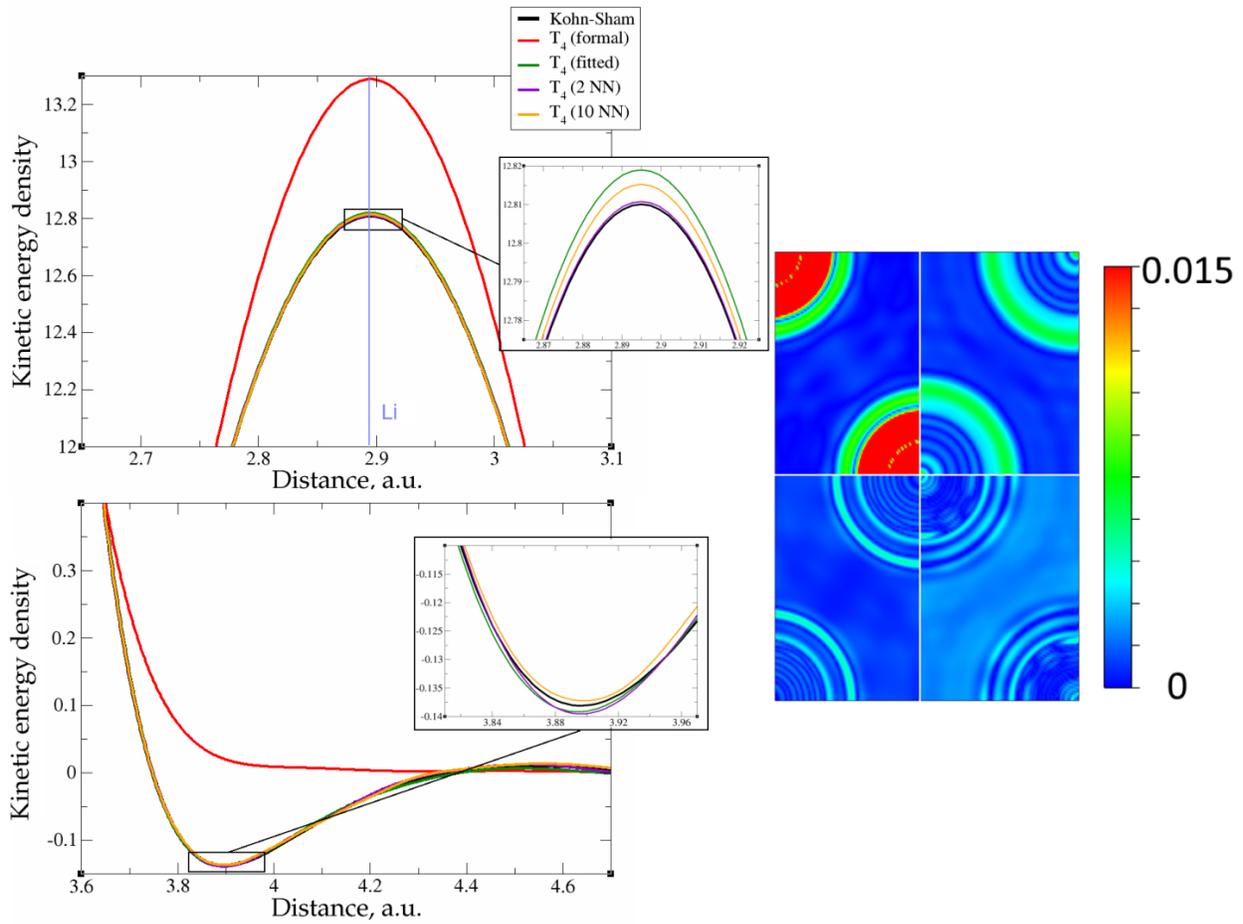

Figure 13. Kinetic energy density from non-linear (NN) fitting for lithium (3 valence electrons). On the left – the line along direction {1 1 1}; on the right – the projection of plane {1 0 1} with absolute differences between KEDs of different gradient approximations and Kohn-Sham KED: upper left quadrant – formal $T^{(4)}$, upper right quadrant – fitted $T^{(4)}$, bottom left quadrant – 2 neurons $T^{(4)}$, bottom right quadrant – 10 neurons $T^{(4)}$.

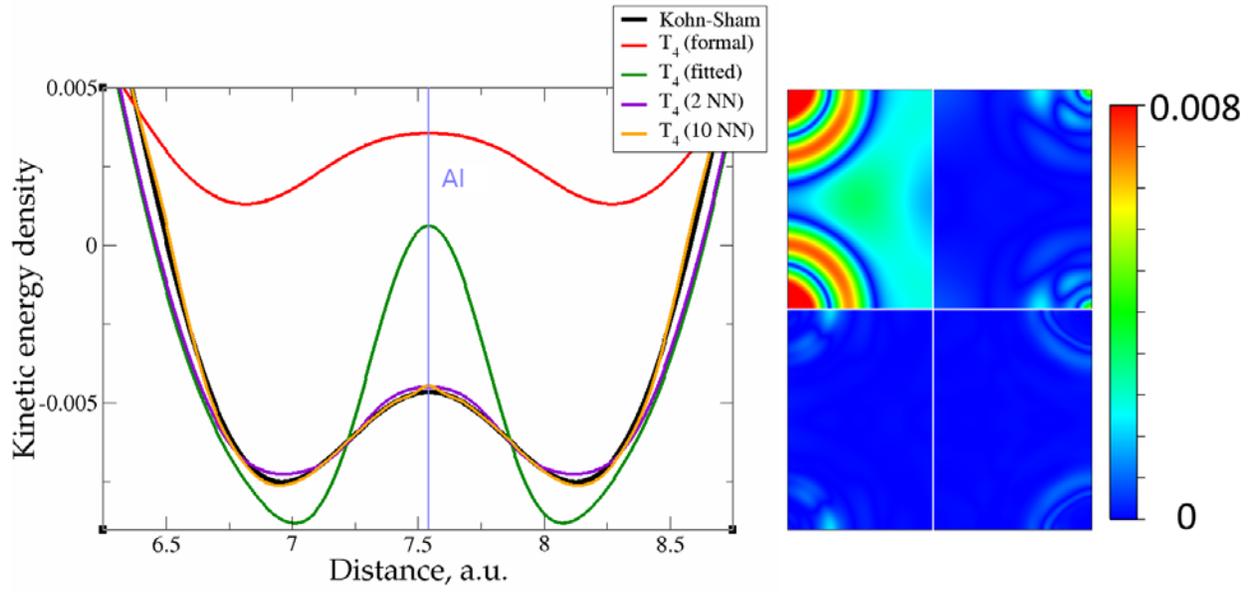

Figure 14. Kinetic energy density from non-linear (NN) fitting for aluminum. On the left – the line along direction {1 0 0}; on the right – the projection of plane {1 0 1} with absolute differences between KEDs of different gradient approximations and Kohn-Sham KED: upper left quadrant – formal $T^{(4)}$, upper right quadrant – fitted $T^{(4)}$, bottom left quadrant – 2 neurons $T^{(4)}$, bottom right quadrant – 10 neurons $T^{(4)}$.

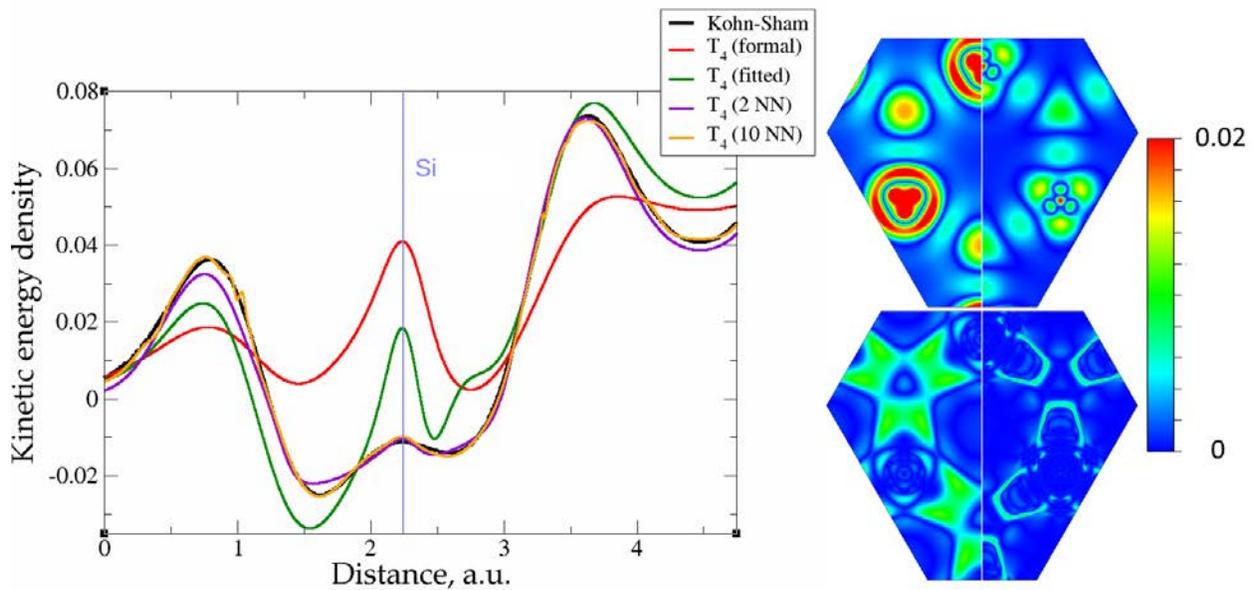

Figure 15. Kinetic energy density from non-linear (NN) fitting for silicon. On the left – the line along direction {1 1 1}; on the right – the projection of plane {1 1 1} with absolute differences between KEDs of different gradient approximations and Kohn-Sham KED: upper left quadrant – formal $T^{(4)}$, upper right quadrant – fitted $T^{(4)}$, bottom left quadrant – 2 neurons $T^{(4)}$, bottom right quadrant – 10 neurons $T^{(4)}$.

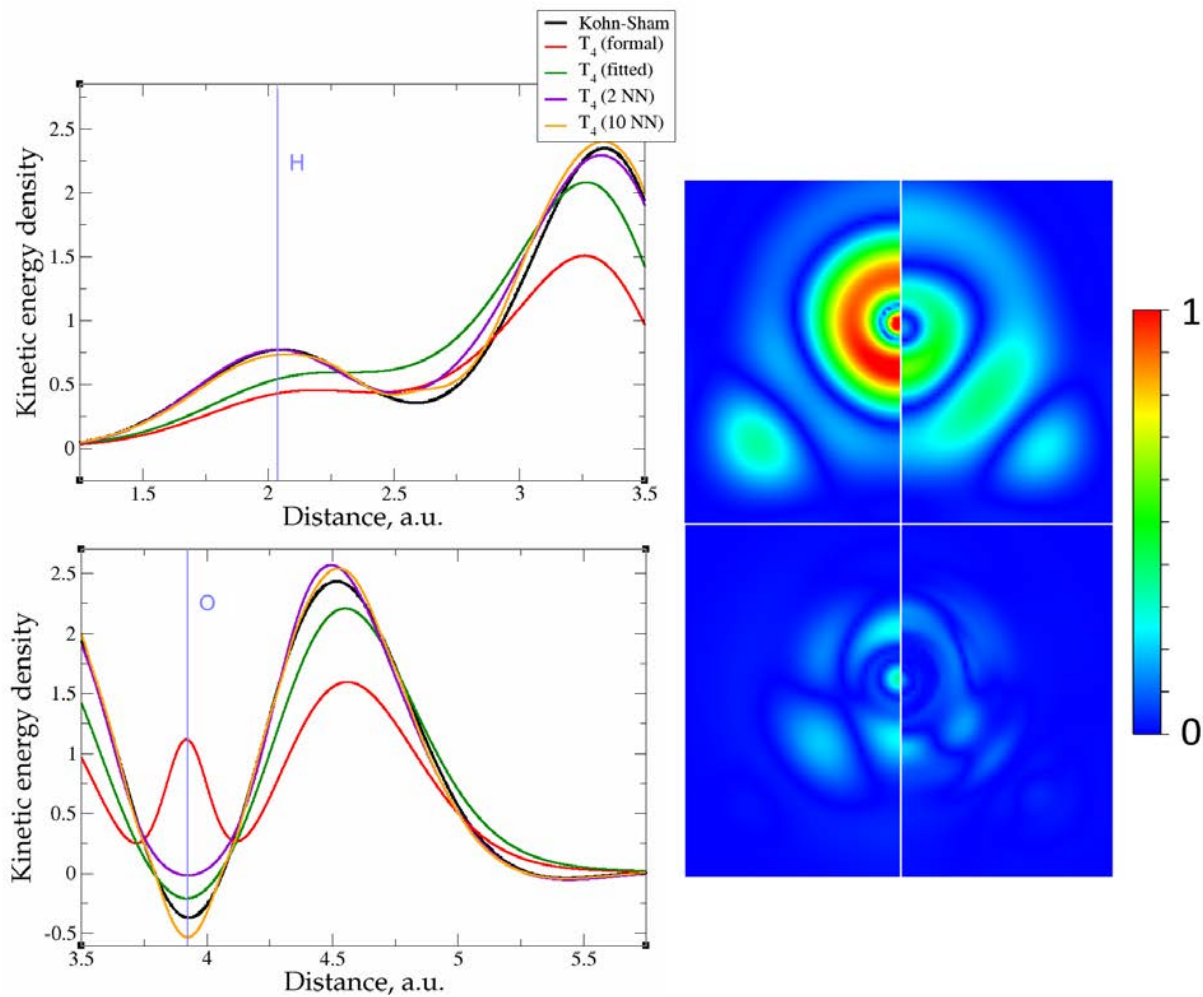

Figure 16. Kinetic energy density from non-linear (NN) fitting for water molecule. On the left – the line along O-H bond; on the right – the projection of molecular plane with absolute differences between KEDs of different gradient approximations and Kohn-Sham KED: upper left quadrant – formal $T^{(4)}$, upper right quadrant – fitted $T^{(4)}$, bottom left quadrant – 2 neurons $T^{(4)}$, bottom right quadrant – 10 neurons $T^{(4)}$.

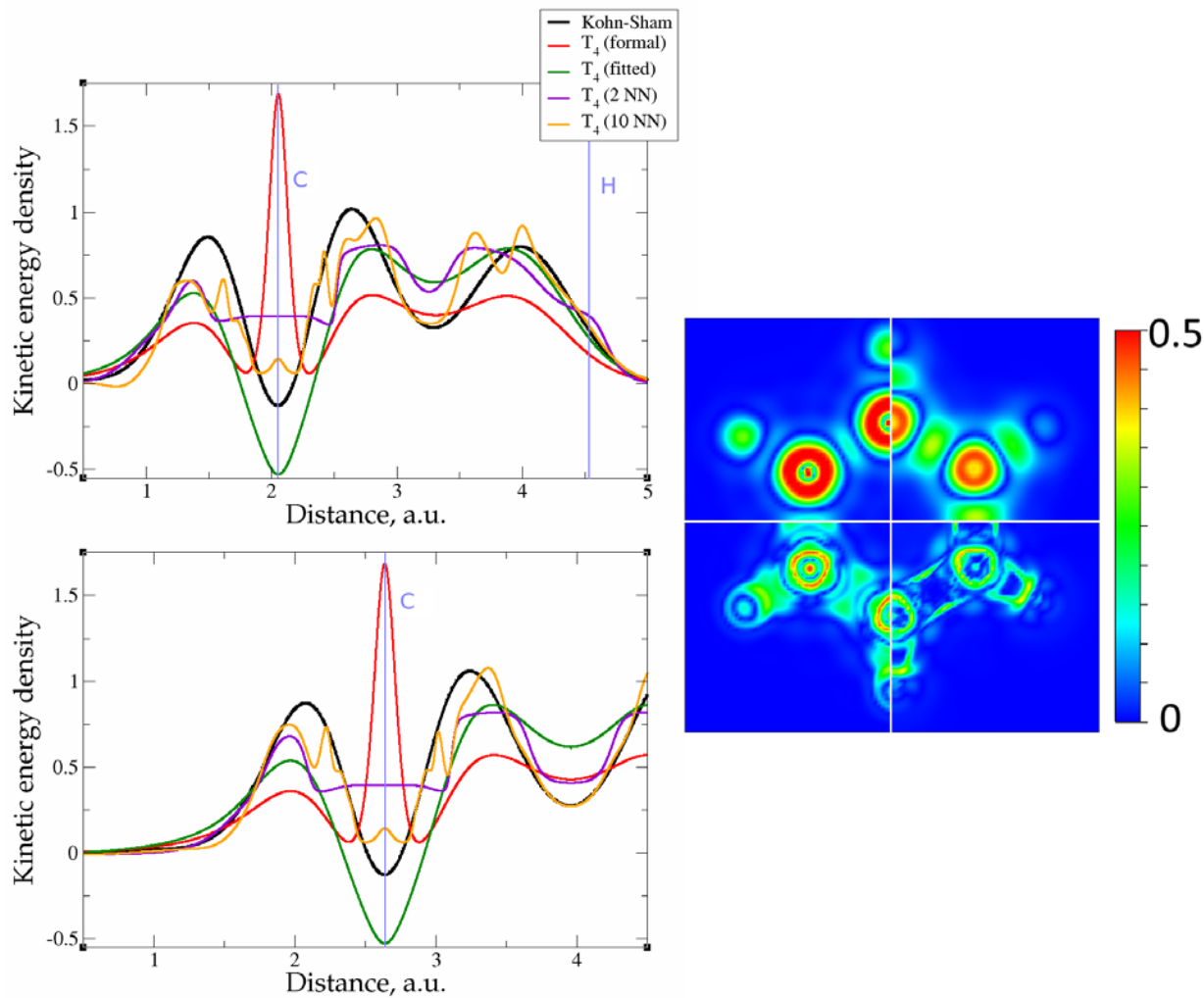

Figure 17. Kinetic energy density from non-linear (NN) fitting for benzene. On the left – the lines along C-H and C-C bonds; on the right – the projection of molecular plane with absolute differences between KEDs of different gradient approximations and Kohn-Sham KED: upper left quadrant – formal $T^{(4)}$, upper right quadrant – fitted $T^{(4)}$, bottom left quadrant – 2 neurons $T^{(4)}$, bottom right quadrant – 10 neurons $T^{(4)}$.

## Joint NN training on KEDs of several compounds

The task of producing a transferable KE functional implies the necessity for performing simultaneous trainings over multiple species of interest (further this procedure will be referred as "joint training"). Since the accuracy of individual trainings for molecules was considerably worse that the accuracy achieved with the solid compounds (although not worse than the ML

KED functionals which were recently reported[52]) we decided to perform joint trainings for Al, Li, Mg and Si. Further, our aim was to reach the accuracy of joint trainings which is comparable with the individual non-linear fits (thus of $10^{-5}$ Ha/Bohr$^3$ for simple metals and $10^{-4}$ Ha/Bohr$^3$ for Si, see Table 3). From this point of view the last are playing role as the reference for a reachable accuracy.

As has been shown above the difficulty of the fit is compounded by the data distribution. While for the individual NN fits the considered above 3D samplings were dense enough, the joint trainings require much denser meshes and seemingly larger neural networks that can severely increase the amount of requisite computational resources. One way to walk around this problem is to make use of the symmetry of the unit cells, i.e. to replace 3D data with 2D data, which can be extracted from the corresponding crystallographic planes. Obviously, the properties (like the electron density and KED) can be sampled much denser over planes than over 3D cells, while the total amount of sampling points will be much less. In our investigation we decided to include into training sets two planes from each of the solid compounds – thus {100} and {110} planes were chosen in the case of Al, Li and Si, {0001} and {11-20} planes were chosen for Mg. The sampling density of the planes was 0.01 Bohr per point and 1/5 of the total amount of points were taken as training data.

The next important point is data compatibility – it appeared that mixing of large-core and small-core elements can make the training procedure much more error prone and more inclined to overfitting. Therefore, we decided to take pseudopotentials with 2 valence electrons for Mg and 1 valence electron for Li. Keeping in mind the pseudopotential issue that was described in section "The role of pseudopotentials", we have taken the corresponding FHI large-core pseudopotentials, which are seemingly less affected by it than the GTH pseudopotentials.

The resulting RMSE's for several neural networks are presented in Table 4. We would like to stress out that the corresponding RMSE's were calculated with respect to the 3D data in the cube files, which were described in the "Introduction" (i.e. beyond the 2D planes which were used for NN training).

Table 4. Root mean square error (Ha/Bohr$^3$) of kinetic energy density integrated over 3D test data for neural network joint fits with different numbers of neurons [$N_n$ ...]. Terms on the 4$^{th}$ order expansion are used as inputs. The amount of numbers inside parentheses ( [$N_n$ ...]) is equal to the number of hidden layers.

| [$N_n$] | Al | Mg$^a$ | Li$^b$ | Si |
|---|---|---|---|---|
| [60] | 3.17×10$^{-4}$ | 2.63×10$^{-4}$ | 1.56×10$^{-4}$ | 1.04×10$^{-3}$ |
| [30 30] | 1.28×10$^{-4}$ | 2.10×10$^{-4}$ | 1.37×10$^{-4}$ | 1.16×10$^{-3}$ |
| [20 20 20] | 9.23×10$^{-5}$ | 1.91×10$^{-4}$ | 2.42×10$^{-5}$ | 1.04×10$^{-3}$ |
| [80] | 2.87×10$^{-4}$ | 1.70×10$^{-4}$ | 6.33×10$^{-5}$ | 9.94×10$^{-4}$ |
| [40 40] | 1.10×10$^{-4}$ | 1.56×10$^{-4}$ | 2.05×10$^{-5}$ | 1.02×10$^{-3}$ |
| [20 20 20 20] | 8.00×10$^{-5}$ | 2.45×10$^{-4}$ | 1.82×10$^{-5}$ | 1.32×10$^{-3}$ |
| [100] | 2.81×10$^{-4}$ | 9.71×10$^{-5}$ | 5.88×10$^{-5}$ | 1.01×10$^{-3}$ |
| [50 50] | 1.16×10$^{-4}$ | 4.90×10$^{-5}$ | 2.21×10$^{-5}$ | 1.14×10$^{-3}$ |
| [25 25 25 25] | 7.56×10$^{-5}$ | 6.99×10$^{-5}$ | 1.82×10$^{-5}$ | 1.69×10$^{-3}$ |
| [20 20 20 20 20] | 3.90×10$^{-5}$ | 2.26×10$^{-5}$ | 1.54×10$^{-5}$ | 1.65×10$^{-3}$ |

$^a$ 2 valence electrons; $^b$ 1 valence electron

It is seen that to get an accuracy which is comparable to the accuracy of the individual fits, one should take 60 or more neurons. As expected, the increase of number of neurons within one hidden layer (thus going from [60] to [80] and [100]) can improve the average (over compounds and data) fit accuracy. This average however hides the fact that the KED may deviate substantially from the target and or exhibit oscillatory behavior is parts of the space, see Figure 18, that makes the resulting KED functionals not suitable for self-consistent density minimization procedures. This is an indication of overfitting caused by data distribution (as the number of NN parameters is much smaller than the number of training data). This can be palliated by using multiple hidden layers. Indeed, the NN configuration [20 20 20] (that is in total 60 neurons, which are distributed evenly over 3 hidden layers) shows the same average accuracy as the NN with configuration [100] (that is, in total 100 neurons in one hidden layer). In general, for all considered total numbers of neurons we find that 2 hidden layers are more advantageous over one hidden layer, and 3 or 4 hidden layers are more advantageous than 2 hidden layers. As shown in Figure 18, the KED functionals which were produced with 4 hidden layers look much smoother. Even when the respective RMSE is not improved, the advantage of several hidden layers is apparent – see as example Si. The RMSE for 80 neurons in one hidden layer is $9.94\times10^{-4}$ a.u. while the RMSE for 80 neurons in 4 hidden layers ([20 20 20 20] in table 4) is slightly larger ($1.32\times10^{-3}$ a.u.); however, the local behavior in the latter case is much better.

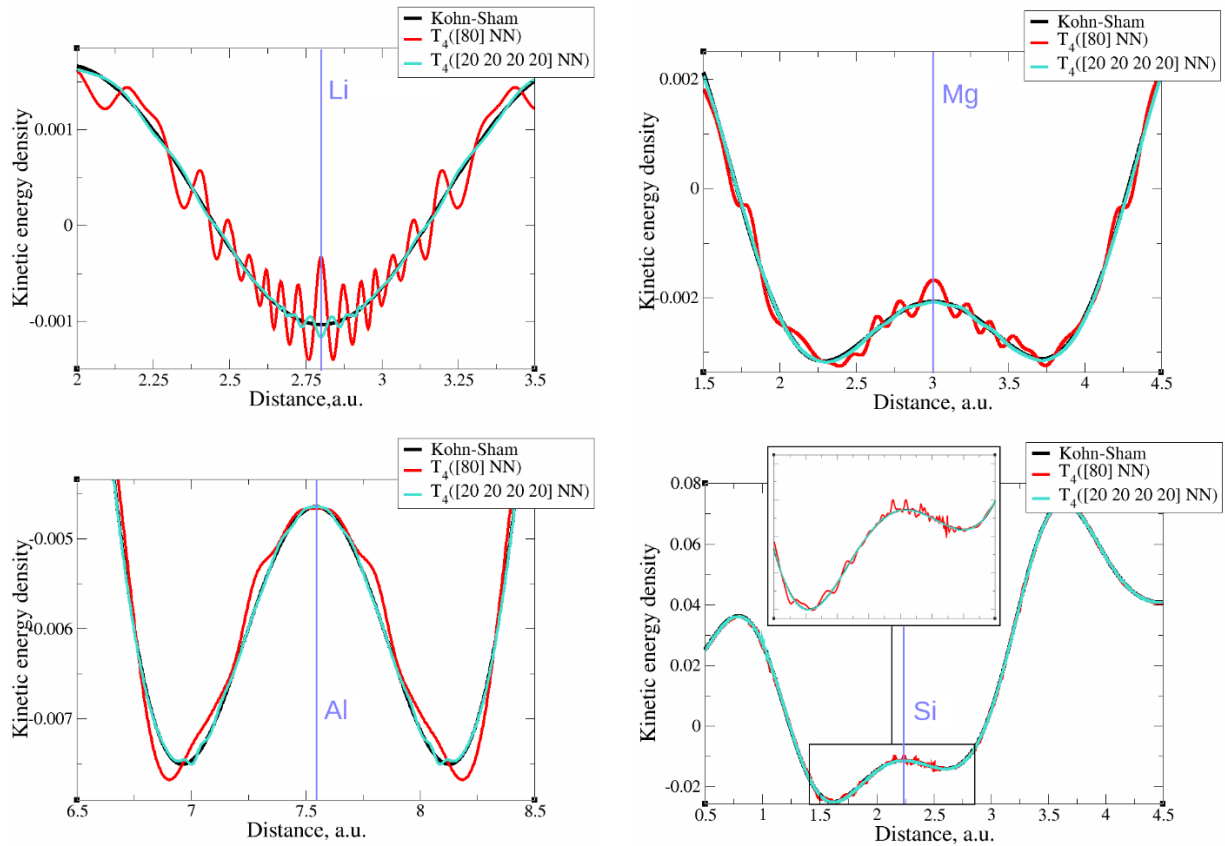

Figure 18. Kinetic energy density from non-linear (NN) fitting of $T^{(4)}$ for: top left - Li, top right - Mg, bottom left - Al, bottom right - Si.

In principle, it must be possible to produce accurate KEDF using this approach for a larger pool of compounds  We would like to emphasize several points that can greatly (as was shown above) facilitate the accuracy of such machine-learned functionals:

- It is necessary to pay a close attention to the data distribution and data selection for training. The KED shows very uneven distribution for all compounds. Data selection can be improved by some effective histogram equalization procedure or by sampling the respective data as close as possible.

- To reduce the computational costs, a sampling over crystallographic planes can be used instead of 3D sampling.

- The utilization of several hidden layers (ideally more than 2) looks important to prevent overfitting and to produce smooth KED in all space.

- There is no need to use extremely large neural networks, 15-20 neurons per compound should be enough to reach a decent accuracy.

**Energy evolution in self-consistent minimization procedure**

In order to verify the performance of $T^{(4)}$ in a self-consistent minimization procedure in OF-DFT, the formal $T^{(4)}$ expansion was implemented in PROFESS 3.0. The electron densities from each SCF optimization step in a Kohn-Sham minimization procedure (done in Abinit) were taken as inputs, and the respective total energies with $T^{(2)}$, $T^{(4)}$, and Wang-Teter (WT)[79] KEDF's were calculated. The results for aluminum and silicon are present in Figure 19 and Figure 20, respectively. Molecules were not considered in this test due to lack of available local pseudopotentials for hydrogen, carbon and oxygen to perform the calculation in PROFESS and because the $T^{(4)}$ approximation performed too poorly for molecules, as shown above.

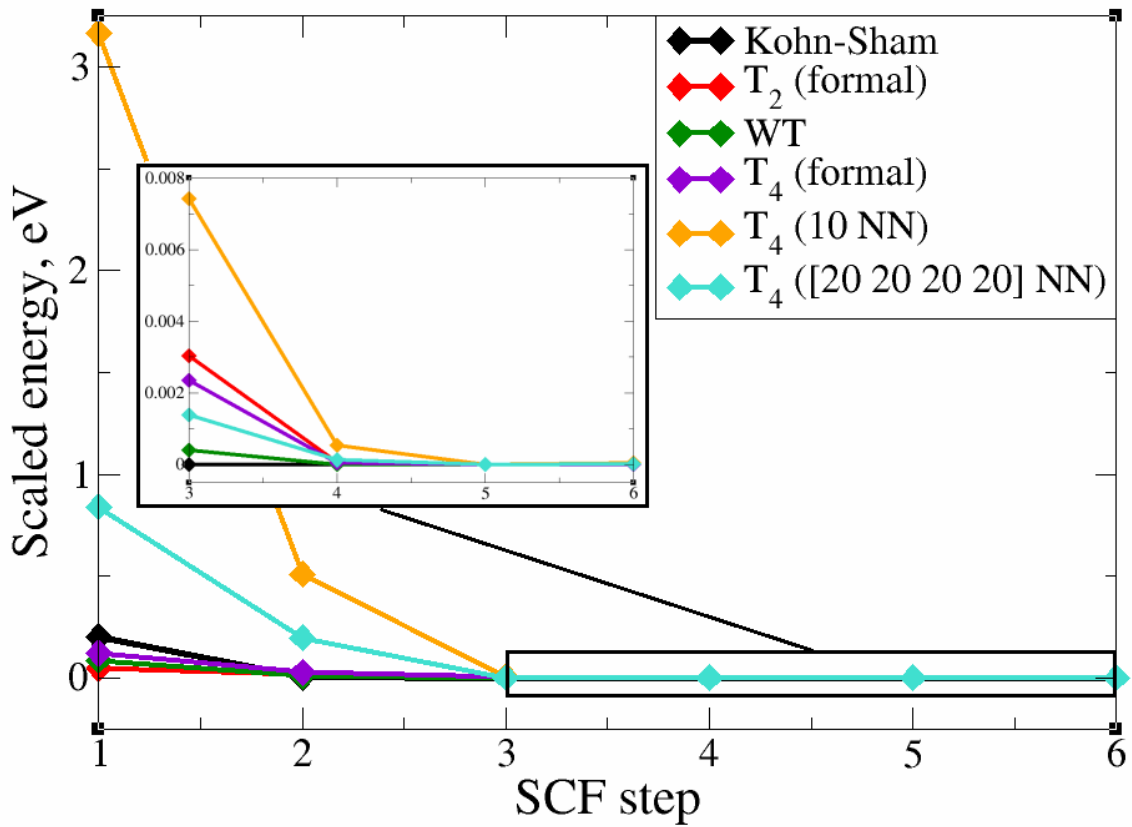

Figure 19. The change of DFT (Abinit) and OF-DFT (PROFESS) total energies for aluminum with input densities from a Kohn-Sham SCF procedure. The energies are plotted with respect to the energy that corresponds to the minimum electronic configuration in given set.

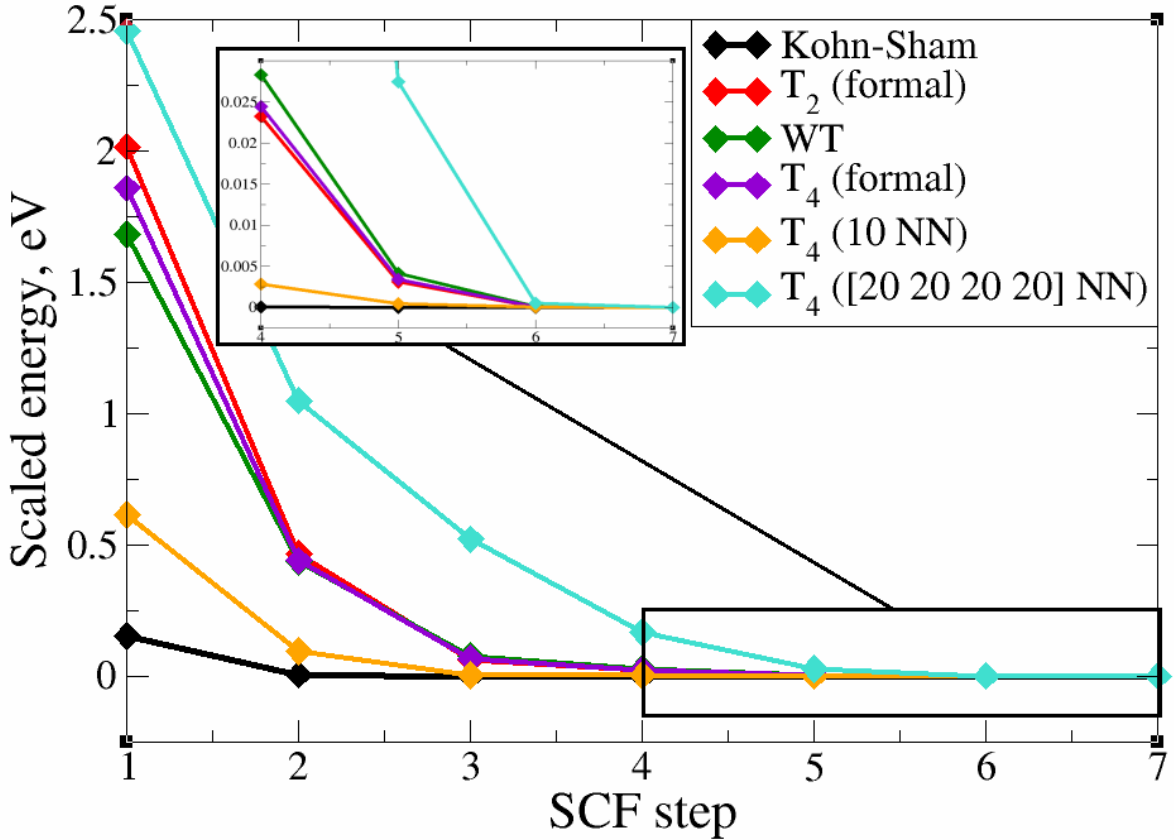

Figure 20. The change of DFT (Abinit) and OF-DFT (PROFESS) total energies for silicon with input densities from a Kohn-Sham SCF procedure. The energies are plotted with respect to the energy that corresponds to the minimum electronic configuration in given set.

The main conclusion is that the formal $T^{(2)}$ expansion, the formal $T^{(4)}$ expansion, the expansions with the fitted coefficients, and non-linear NN fits all show a monotonic decrease of energy along KS-DFT SCF steps, as does the WT KEDF, even in the case of Si. This is encouraging. However, we were unable to perform a complete SCF minimization neither with the formal $T^{(4)}$ expansion nor with the fits. Depending on different input electron densities (whether uniform, which is the default initialization in PROFESS, or taken from Kohn-Sham calculation), the optimization procedures ended up stuck in relatively deep local minima. It is however possible that this error happened due to local pseudopotentials in OF-DFT calculations, since the numerical instabilities in our work were caused exclusively by pseudopotentials, and the pseudopotentials for Al and Si available in PROFESS do have a shape (a "wall" towards the nucleus, see Figure 11) which we showed above cause instabilities in the terms of the gradient expansion.

## Conclusions

We explored the possibility of using the 4th order expansions of electron density as a base for the search of orbital-free approximations to KED. It was previously suggested that due to the electron densities of different powers in denominators, these expansions would be numerically unstable. However, we have shown that with the proper choice of pseudopotential, the corresponding KED curves from 4th-order expansions remain continuous everywhere, including low-density regions. For simple bulk metals Li, Al, Mg, the 4th order expansion of the electron density with fitted coefficients is already able to reproduce the Kohn-Sham KED well. This is not the case for covalently bound systems such as bulk Si and molecular $H_2O$ and $C_6H_6$ tried here.

A closer match to the Kohn-Sham KED, including for systems with localized electrons (covalent bonds), can be obtained by using the 4th order expansion terms as inputs for non-linear regression, which we performed using neural networks. Our NN fits to each compound resulted in extremely low fit and test set RMSE values without overfitting of NN coefficients. For this purpose, small single layer NNs serve well, since they are already universal approximators.

For covalently bonded compounds, the quality of the 4th order expansion is strongly affected by the non-uniformity of KED distributions. A better representability of the fitting data could be achieved by consideration of problematic regions (generally near-nuclei regions) on denser grids. More advanced techniques to deal with data distribution are worth to trying for.[80]

The evolution of total energies from $T^{(4)}$ expansions, calculated with Kohn-Sham electron densities as inputs, is seemingly in accordance with Kohn-Sham energies – the minimum of the electron density corresponds to minimum of the total energy. However, the elaboration of better local pseudopotentials might be required in order to use $T^{(4)}$ expansions or its parts in SCF minimization procedures, which in our tests were not convergent. We thus hope that these results will be useful in the development of new kinetic energy functionals and pseudopotentials for orbital-free DFT.

Further, the issue of non-uniformity of KED distributions greatly complicates NN trainings to KEDs of several compounds simultaneously, and, as consequence, the task of producing of reliable transferable KE functional. We have shown that good joint fits can be obtained by densely sampling training data on planes in combination with the usage of several hidden layers in NN architecture. We would like to point out that this does not question the quality of a single hidden layer network as universal approximator, since this holds only for infinitely dense sampling.

The main result is that it is possible to obtain very close match to orbital-based KEDs using machine learning tools, both for individual compounds (including molecules) and joint trainings. We achieved this with the terms of $T^{(4)}$ expansion as density-dependent inputs. The corresponding RMSEs were extremely low, in order of 1 mHa/Bohr$^3$ (for Si) and less (for simple metals). To the best of our knowledge, it is the most precise NN fitting to KEDs of real systems so far.

## Acknowledgements

This work was supported by the Ministry of Education of Singapore (AcRF Tier 1 grant). We thank Daniel Koch for help with some calculations and Michele Pavanello, Emily Carter, Paul Ayers, and Hiromi Nakai for discussions.

## References

1. P. Hohenberg and W. Kohn, *Phys. Rev.*, 1964, **136**, B864.
2. D. Sholl and J. A. Steckel, *Density Functional Theory: A Practical Introduction,* Wiley, 2009.
3. W. Kohn and L. J. Sham, *Phys. Rev.*, 1965, **140**, A1133.

4. T. A. Wesolowski and Y. A. Wang (Eds.), *Recent Progress in Orbital-free Density Functional Theory,* World Scientific, 2013.

5. Y. A. Wang, N. Govind and E. A. Carter, *Phys. Rev. B,* 1998, **58**, 13465.

6. Y. A. Wang, N. Govind and E. A. Carter, *Phys. Rev. B,* 1999, **60**, 16350.

7. B. Radhakrishnan and V. Gavini, *Philosophical Magazine,* 2016, **96**, 2468.

8. S. Das, M. Iyer and V. Gavini, *Phys Rev. B,* 2015, **92**, 014104.

9. M. Chen, X.-W. Jiang, H. Zhuang, L.-W. Wang and E. A. Carter, *J. Chem. Theory Comput*., 2016, **12**, 2950.

10. W. C. Witt, B. G. del Rio, J. M. Dieterich and E. A. Carter, *J. Mater. Res.,* 2018, **33**, 777.

11. C. Huang and E. A. Carter, *Phys. Rev. B,* 2010, **81**, 045206.

12. D. A. Kirzhnits, *Sov. Phys.-JETP,* 1957, **5**, 64.

13. L. H. Thomas, *Proc. Cambridge Phil. Soc*., 1927, **23**, 542.

14. E. Fermi, *Rend. Accad. Naz. Lincei*, 1927, **6**, 602.

15. C. F. V. Weizsäcker, *Z. Phys*., 1935, **96**, 431.

16. C. H. Hodges, *Can. J. Phys.,* 1973, **51**, 1428.

17. D. R. Murphy, *Phys. Rev. A,* 1981, **24**, 1682.

18. W.-P. Wang, R. G. Parr, D. R. Murphy and G. A. Henderson, *Chem. Phys. Lett*., 1976, **43**, 409.

19. D. R. Murphy and W.-P. Wang, *J. Chem. Phys*., 1980, **72**, 429.

20. E. Clementi and C. Roetti, *At. Data Nucl. Data Tables,* 1974, **14**, 177.

21. C. Lee and S. K. Ghosh, *Phys. Rev. A,* 1986, **33**, 3506.

22. N. L. Allan, C. G. West, D. L. Cooper, P. J. Grout and N. H. March, *J. Chem. Phys*., 1985, **83**, 4562.

23. N. L. Allan and D. L. Cooper, *J. Chem. Phys*., 1986, **84**, 5594.

24. P. E. Blöchl, *Phys. Rev. B,* 1994, **54**, 17953.

25. J. Lehtomäki, I. Makkonen, M. A. Caro, A. Harju and O. Lopez-Acevedo, *J. Chem. Phys.* 2014, **141**, 234102.

26. Y. Tal and R. F. W. Bader, *Int. J. Quantum Chem*., 1978, **S12**, 153.

27. E. W. Pearson and R. G. Gordon, *J. Chem. Phys*., 1985, **82**, 881.

28. J. P. Perdew, M. Levy, G. S. Painter, S. Wei and J. B. Lagowski, *Phys. Rev. B,* 1988, **37**, 838.

29. A. E. DePristo and J. D. Kress, *Phys. Rev. A,* 1987, **35**, 438.

30. A. Sergeev, F. H. Alhabri, R. Jovanovic and S. Kais, *J. Phys.: Conf. Ser*., 2016, **707**, 012011.

31. Z. Yan, J. P. Perdew, T. Korhonen and P. Ziesche, *Phys. Rev. A,* 1997, **55**, 4601.

32. L. Vitos, H. L. Skriver and J. Kollár, *Phys. Rev. B,* 1998, **57**, 12611.
33. S. Laricchia, L. A. Constantin, E. Fabiano and F. Della Sala, *J. Chem. Theory Comput.,* 2014, **10(1)**, 164.
34. L. A. Constantin, E. Fabiano and F. Della Sala, *J. Phys. Chem. Lett.,* 2018, **9**, 4385.
35. J. Tao, J. P. Perdew, V. N. Staroverov and G. E. Scuseria, *Phys. Rev. Lett.,* 2003, **91**, 146401.
36. J. P. Perdew and L. A. Constantin, *Phys. Rev. B,* 2007, **75**, 155109.
37. M. Levy and H. Ou-Yang, *Phys. Rev. A,* 1988, **38**, 625.
38. S. B. Trickey, V. V. Karasiev and R. S. Jones, *Int. J. Quantum Chem.* 2009, **109**, 2943.
39. V. V. Karasiev, D. Chakrabotry, O. A. Shukruto and S. B. Trickey, *Phys. Rev. B,* 2013, **88**, 161108(R).
40. S. Smiga, E. Fabiano, L. A. Constantin and F. Della Sala, *J. Chem. Phys.,* 2017, **146**, 064105.
41. J. Hollingsworth, L. Li, T. E. Baker and K. Burke, *J. Chem. Phys.,* 2018, **148**, 241743.
42. T. M. Mitchell, *Machine Learning,* MacGraw-Hill Science, 1997.
43. S. Manzhos, X.-G. Wang, R. Dawes and T. A. Carrington, *J. Phys. Chem. A,* 2006, **110**, 5295.
44. S. Manzhos, R. Dawes and T. Carrington, *Int. J. Quantum Chem.*, 2015, **115**, 1012.
45. B. Kolb, P. Marshall, B. Zhao, B. Jiang and H. Guo, *J. Phys. Chem. A,* 2017, **121**, 2552.
46. G. B. Goh, N. O. Hodas and A. Vishnu, *J. Comput. Chem.*, 2017, **38**, 1291.
47. J. C. Snyder, M. Rupp, K. Hansen, L. Blooston, K.-R. Müller and K. Burke, *J. Chem. Phys*., 2013, **139**, 224104.
48. L. Li, J. C. Snyder, I. M. Pelaschier, J. Huang, U.-M. Niranjan, P. Duncan, M. Rupp, K.-R. Müller and K. Burke, *Int. J. Quantum Chem.*, 2016, **116**, 819.
49. L. Li, T. E. Baker, S. R. White and K. Burke, *Phys. Rev. B,* 2016, **94**, 245129.
50. K. Yao and J. Parkhill, *J. Chem. Theory Comput.*, 2016, **12**, 1139.
51. F. Brockherde, L. Vogt, L. Li, M. E. Tuckerman, K. Burke and K.-R. Müller, *Nature Commun.*, 2017, **8**, 872.
52. J. Seino, R. Kageyama, M. Fujinami, Y. Ikabata and H. Nakai, *J. Chem. Phys.,* 2018, **148**, 241705.
53. X. Gonze, J.-M. Beuken, R. Caracas, F. Detraux, M. Fuchs, G.-M. Rignanese, L. Sindic, M. Verstraete, G. Zerah, F. Jollet, M. Torrent, A. Roy, M. Mikami, P. Ghosez, J.-Y. Raty and D. C. Allan, *Comput. Mater. Sci.,* 2002, **25**, 478.
54. X. Gonze, B. Amadon, P.-M. Angade, J.-M. Beuken, F. Bottin, P. Boulanger, F. Bruneval, D. Caliste, R. Caracas, M. Cote, T. Deutsch, L. Genovese, P. Ghosez, M. Giantomassi, S. Goedecker, D. R. Hamann, P. Hermet, F. Jollet, G. Jomard, S. Leroux, M. Mancini, S. Mazevet, M. J. T. Oliveira, G. Onida, Y. Pouillon, T. Rangel, G.-M. Rignanese, D. Sangalli,

R. Shaltaf, M. Torrent, M. J. Verstraete, G. Zerah and J. W. Zwanziger, *Comput. Phys. Commun.*, 2009, **180**, 2582.

55. J. P. Perdew, K. Burke and M. Ernzerhof, *Phys. Rev. Lett.*, 1996, **77**, 3865.
56. W. Witt, *Z. Naturforsch. A,* 1967, **22A**, 92; M. R. Nadler and C. P. Kempfer, *Anal. Chem.,* 1959, **31**, 2109; C. B. Walker and M. Marezio, *Acta Metallurgica,* 1959, **7**, 769; C.R. Hubbard, H. E. Swanson and F. A. Mauer, *J. Appl. Crystallogr.*, 1975, **8**, 45.
57. S. Goedecker, M. Teter and J. Hutter, *Phys. Rev. B,* 1996, **54**, 1703.
58. M. Krack, *Theor. Chem. Acc.,* 2005, **114**, 145.
59. M. Fuchs and M. Scheffler, *Comput. Phys. Commun.,* 1999, **119**, 67.
60. Pseudopotentials for the ABINIT code, https://www.abinit.org/sites/default/files/PrevAtomicData/psp-links/psp-links/gga_fhi.
61. Neural Network Toolbox, https://www.mathworks.com/products/neural-network.html.
62. Gaussian cube files, http://paulbourke.net/dataformats/cube/.
63. R. Gonzalez and R. Woods, *Digital Image Processing*, 2nd Ed., New Jersey: Prentice Hall, 2002.
64. G. S. Ho, V. L. Ligneres and E. A. Carter, *Comput. Phys. Commun.*, 2008, **179**, 839.
65. V. V. Karasiev and S. B. Trickey, *Comput. Phys. Commun.*, 2012, **183**, 2519.
66. W. Mi, S. Zhang, Y. Wang, Y. Ma and M. Miao, *J. Chem. Phys.,* 2016, **144**, 134108.
67. C. Huang and E. A. Carter, *Phys. Chem. Chem. Phys.,* 2008, **10**, 7109.
68. F. Tran and T. A. Wesolowski, *Int. J. Quantum Chem.,* 2002, **89**, 441.
69. V. V. Karasiev, S. B. Trickey and F. E. Harris, *J. Comput.-Aided Mater. Des.,* 2006, **13**, 111.
70. V. V. Karasiev and S. B. Trickey, *Advances in Quantum Chem.,* 2015, **71**, 221.
71. K. Luo, V. V. Karasiev and S. B. Trickey, *Phys. Rev. B,* 2018, **98**, 041111(R).
72. K. Hornik, M. Stinchcombe and H. White, *Neural Networks,* 1989, **2**, 359.
73. K. Hornik, *Neural Networks,* 1991, **4**, 251.
74. J. Schwinger, *Phys. Rev. A,* 1980, **22**, 1827.
75. P. K. Acharya, L. J. Bartolotti, S. B. Sears and R. G. Parr, *Proc Natl. Acad. Sci. USA,* 1980, **77**, 6978.
76. W. Yang, *Phys. Rev. A,* 1986, **34**, 4575.
77. G. K.-L. Chan, A. J. Cohen and N. C. Handy, *J. Chem. Phys.,* 2001, **114**, 631.
78. L. A. Espinosa Leal, A. Karpenko, M. A. Caro and O. Lopez-Acevedo, *Phys. Chem. Chem. Phys.,* 2015, **17**, 31463.
79. L.-W. Wang and M. P. Teter, *Phys. Rev. B,* 1992, **45**, 13196.
80. J. Xia and E. A. Carter, *Phys Rev. B,* 2015, **91**, 045124.